\documentclass[sigconf,natbib=true,nonacm]{acmart}

\usepackage{float}

\usepackage{amsfonts,amssymb} 
\usepackage{algorithm}  
\usepackage{algorithmic}  

\usepackage{multirow}
\usepackage{booktabs}

\usepackage{makecell}

\usepackage{enumitem}

\usepackage{array} 

\usepackage{setspace}

\AtBeginDocument{%
  \providecommand\BibTeX{{%
    \normalfont B\kern-0.5em{\scshape i\kern-0.25em b}\kern-0.8em\TeX}}}

\begin{document}

\title{Diffusion Cross-domain Recommendation}

% \affiliation{%
%   \country{cikm}

% \affiliation{%
% \affiliation{%

\author{Yuner Xuan}
\affiliation{%
  \institution{Chalmers University of Technology}
  \institution{Computer Science and Engineering Department}
  \streetaddress{SE-412 96}
  \city{Gothenburg}
  \country{Sweden}}
\email{yuner@chalmers.se}

\renewcommand{\shortauthors}{Xuan}

\begin{abstract}
It is always a challenge for recommender systems to give high-quality outcomes to cold-start users.
One potential solution to alleviate the data sparsity problem for cold-start users in the target domain is to add data from the auxiliary domain.
Finding a proper way to extract knowledge from an auxiliary domain and transfer it into a target domain is one of the main objectives for cross-domain recommendation (CDR) research. 
Among the existing methods, mapping approach is a popular one to implement cross-domain recommendation models (CDRs). 
For models of this type, a mapping module plays the role of transforming data from one domain to another.
It primarily determines the performance of mapping approach CDRs.
Recently, diffusion probability models (DPMs) have achieved impressive success for image synthesis related tasks. They involve recovering images from noise-added samples, which can  be viewed as a data transformation process with outstanding performance. 
To further enhance the performance of CDRs, we first reveal the potential connection between DPMs and mapping modules of CDRs, and then propose a novel CDR model named Diffusion Cross-domain Recommendation (DiffCDR)\footnote{ \url{https://github.com/breezeyuner/DiffCDR}}.
More specifically, we first adopt the theory of DPM and design a Diffusion Module (DIM), which generates user's embedding in target domain.  
To reduce the negative impact of randomness introduced in DIM and improve the stability, we employ an  Alignment Module to produce the aligned user embeddings. In addition, we consider the label data of the target domain and form the task-oriented loss function, which enables our DiffCDR to adapt to specific tasks. By conducting extensive experiments on datasets collected from reality, we  demonstrate the effectiveness and adaptability of DiffCDR to outperform baseline models on various CDR tasks in both cold-start and warm-start scenarios.
\end{abstract}

\begin{CCSXML}
<ccs2012>
 <concept>
  <concept_id>10010520.10010553.10010562</concept_id>
  <concept_desc>Computer systems organization~Embedded systems</concept_desc>
  <concept_significance>500</concept_significance>
 </concept>
 <concept>
  <concept_id>10010520.10010575.10010755</concept_id>
  <concept_desc>Computer systems organization~Redundancy</concept_desc>
  <concept_significance>300</concept_significance>
 </concept>
 <concept>
  <concept_id>10010520.10010553.10010554</concept_id>
  <concept_desc>Computer systems organization~Robotics</concept_desc>
  <concept_significance>100</concept_significance>
 </concept>
 <concept>
  <concept_id>10003033.10003083.10003095</concept_id>
  <concept_desc>Networks~Network reliability</concept_desc>
  <concept_significance>100</concept_significance>
 </concept>
</ccs2012>
\end{CCSXML}

\ccsdesc[500]{Information systems~Recommender systems}

\keywords{Cross-domain Recommendation, Diffusion Probability Model, Cold-start Problem, Mapping Approach}

%% \begin{teaserfigure}
%%   \includegraphics[width=\textwidth]{sampleteaser}
%%   \caption{Seattle Mariners at Spring Training, 2010.}
%%   \Description{Enjoying the baseball game from the third-base
%%   seats. Ichiro Suzuki preparing to bat.}
%%   \label{fig:teaser}
%% \end{teaserfigure}


\maketitle

%------------------------------
\section{Introduction}
%a.
\noindent $\bullet$ \textbf{Introduction of CDR.} Recommender systems (RSs) have become a basis technique in various web and mobile applications, including Amazon (e-Commerce), YouTube (Video Platform), Facebook (Social Network Service) and Booking (Online Travel Agency). This field has attracted a large number of researchers, both from academia and industry.

Nonetheless, RSs still face significant challenges in real-world applications. 
The cold-start user problem is one of the most mentioned challenges. Since a large part of purchasing and rating records are generated by a few users \cite{ricci2011introduction}, it can be difficult for RSs to provide satisfying recommendations for users who have no interaction in the past. These users are usually named cold-start users. Studies on cross-domain recommendation (CDR)  \cite{singh2008relational} aim to deal with this issue.

%b.
The core motivation of CDR is to transfer knowledge from a source domain with richer interaction records of a set of users to a target domain without historical interactions of the same users\cite{zhu2021cross,DBLP:journals/corr/abs-2108-03357}. CDR models (CDRs) can alleviate the cold-start users problem and improve the recommendation results.

%c.
%这里应该只是泛泛介绍，一句话概括多篇文章。
%每一篇具体的介绍，放在related work
Various deep learning models have been proposed in recent years to boost the performance on knowledge transfer. 
Embedding-based mapping approach  \cite{zhu2021cross} is one of widely studied cross-domain knowledge transfer methods.
The first step of this method is training collaborative filtering based models  \cite{koren2009matrix,rendle2012bpr,singh2008relational} to produce user's and item's embeddings in different domains respectively. Second, it transfers these embeddings across domains via various mapping functions. Typically, the mapping function is trained on data from the same users or items in different domains. In the prediction phase, the model maps corresponding embeddings from one domain to another.

%接下来是粗略过一下不同mapping 方法的工作
Existing mapping approach CDRs fall into three main types.
\textbf{1.Mapping function learning:}
Mapping function learning methods  \cite{man2017cross,zhu2020deep,kang2019semi} employ different techniques to learn mapping functions directly, which are used to transfer the knowledge across source and target domains.
They take advantage of the data from overlapping users in different domains and learn the mapping patterns.
\textbf{2.Meta learning:}
Meta learning \cite{munkhdalai2017meta} is popular in recent RS studies \cite{pan2019warm,lee2019melu,zhu2022personalized}.
For the works related to CDR \cite{zhu2022personalized,zhu2021learning}, they adopt meta-learning methods to generate mapping functions and variables to overcome the problem of cold-start.
\textbf{3.Variational Autoencoder (VAE):}
Due to the successful applications of VAE in RSs  \cite{liang2018variational,chen2018collective,shenbin2020recvae}, an increasing number of researchers have proposed VAE-based methods \cite{cao2022cross,salah2021towards,liu2022exploiting} to tackle CDR tasks. These methods align the latent embedding variables of VAEs across domains through regularizers and alignment strategies. By narrowing the distribution divergence of latent variables between domains, they may generate better CDR results both for cold-start and warm-start users.

%d.
%映射方案依然有提升的空间

%%1. corss-domain transfer的质量，决定于 mapping function 的能力。 DM 在密度估计上的优势， （ DM  have achieved state-of-the-art results in density estimation  \cite{kingma2021variational}）能够更好学习target domain上user embedding 的分布，从而提升transfer的 performance

%%2. 存在 noising label data ;在ablation 实验中，从结果看可以表明，若只依赖于 task label learning 并不一定提升效果,

%以上两个原因， 更好的 density estimation，并且独立在 transfer 任务上训练以保证其质量，可以提升CDR 的性能

\noindent $\bullet$ \textbf{Analysis of existing methods. }The objective of the mapping approach CDRs is straightforward and intuitive. The essential challenge for this approach is how to learn mapping patterns effectively \cite{zhu2021cross}.
Although existing works have made significant progress as mentioned above, there are still several issues that deserve attention. 

First, to have a better recommendation performance, CDRs are often optimized on task labels (task-oriented) \cite{zhu2022personalized}, and the quality of the mapping function is treated as an intermediate and auxiliary factor. 
Therefore, the quality of the knowledge transferring could make way for task-specific objectives.
As a consequence, CDRs can easily over-fit on task labels and be deteriorated by the noise existing in label data. According to our ablation experiments, there is no guarantee of achieving better performance by task-oriented learning strategy.

Second, the mapping functions employed in existing mapping-oriented works\cite{man2017cross,kang2019semi,10.1145/3459637.3482137} which are likely to consider the transferring quality for each data sample, rather than narrow the distributional divergence across domains.  It may restrict their generalization ability on unseen samples.

Therefore, if a CDR model is powerful to estimate the distribution of the target embedding space and also considers specific recommendation tasks, it is promising to achieve better recommendation performance.

%e. diffusion 模型简介
\noindent $\bullet$ \textbf{Diffusion probability models (DPMs). }
DPMs \cite{ho2020denoising,dhariwal2021diffusion,rombach2022high,ramesh2022hierarchical} have achieved impressive results in image synthesis recently.
In general, DPMs include forward and backward processes. 
In the forward process, a small amount of noise is gradually added to the original data, which can be overwhelmed by the noise for a sufficient number of steps. Oppositely, the noise will be removed in the reverse process. Since the noise can be simple (e.g. Gaussian noise) with small magnitude in each step of forward process, the denoised data generated in reverse process can achieve satisfying quality. 
One of the obvious deficiencies of diffusion models is the low speed of inference. In the reverse process, it takes hundreds or even thousands of steps from pure random noise to generate a final sample. A lot of works focus on solving this problem  \cite{song2020denoising,bao2022analytic,salimans2022progressive,lu2022dpm}, and they have accelerated the sampling process by dozens of times successfully.

%f. Diffusion based cross-domain recommendation 
% 前面提到的两个潜在改进点，1.overemphasize on tasks 2. noisy labels
% 需要针对性地提一下

\noindent $\bullet$ \textbf{Motivation.} As a state-of-the-art method of density estimation\cite{kingma2021variational} and high-quality sampling for image data, DPMs is able to play the role as a powerful mapping module in CDRs.
DPMs take noisy data as input and generate denoised results, which can be viewed as a process of transferring data from one distribution to another. This property makes the DPMs framework as a promising approach for knowledge transfer in CDR tasks.

\begin{figure}
\centering
\includegraphics[width=0.9\linewidth]{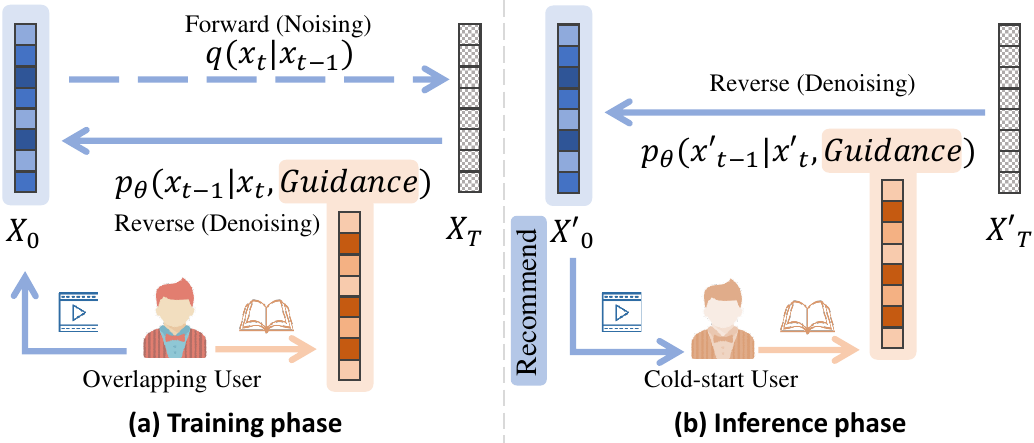}
\caption{Illustration of the DiffCDR concept. (a) Learning the Diffusion Module (DIM) $p_\theta$ on the embeddings of overlapping users (b) Generating transferred embeddings via reverse diffusion process for cold-start users conditioned on corresponding latent vectors from auxiliary domain.} 
\label{fig_concept}
\end{figure}

To enhance the performance of the mapping functions by employing DPMs, we propose an innovative CDR framework named Diffusion Cross-domain Recommendation (DiffCDR) in this paper. 
\textbf{1.} In DiffCDR, we learn a Diffusion Module (DIM) to transfer knowledge across domains. DIM generates user features in the target domain by reversing the diffusion process conditioned on the corresponding user's embeddings in source domain. We illustrate the concept of training and prediction phases in Figure (\ref{fig_concept}). Our experiments results (RQ3) show that DIM mainly contributes to generate better quality transferred users embeddings than baseline models.
\textbf{2.}
As a generative method, randomness may be introduced into a DPM both in training and inference phases. 
To improve the stability of recommendation outcomes, we also design an Alignment Module (ALM). It ensures the consistency between the transformed user's embeddings and ground-truth latent vectors of the same user in target domain.
\textbf{3.} In addition, we also employ a target label data learning strategy to take the final recommendation quality into consideration.

%g. contribution 概括
\noindent $\bullet$ \textbf{Contributions.} We summarize the main contributions of our work as follows: 
\textbf{(1)} In order to improve the performance of mapping approach CDRs, we employ DPM framework and design a Diffusion Module (DIM), which transfers user's preference from interaction data in auxiliary domain. To the best of our knowledge, this is the first work to apply DPM framework to CDR tasks.
\textbf{(2)} To alleviate the negative influence of randomness and ensure the performance of cross-domain knowledge transfer, we employ an Alignment Module (ALM), which takes the transferred embeddings of DIM as input and generates aligned user embedding vectors. To ensure the recommendation performance, DiffCDR employs a task label data learning strategy as well.
\textbf{(3)} For evaluation, we conduct extensive experiments on the Amazon reviews dataset to demonstrate the effectiveness and adaptability of DiffCDR in both cold-start and warm-start CDR scenarios. The ablation experiments show that all components of DiffCDR are critical to contribute to the improvements.

%------------------------------
\section{Background}
\subsection{Problem Setting}
In our CDR tasks, there is a source domain and a target domain. 
In each domain, we have a user set $\mathcal{U}  = \{ u_1,u_2,...  \}$, an item set $\mathcal{V}  = \{ v_1,v_2,...  \}$, and a rating matrix $\mathcal{R}$. For each element of $\mathcal{R}$, $r_{ij}$ denotes the rating by $u_i$ on the item $v_j$.
We use $\mathcal{U}^s$,$\mathcal{V}^s$,$\mathcal{R}^s$ to denote the source domain, and $\mathcal{U}^t$,$\mathcal{V}^t$,$\mathcal{R}^t$ for the target domain.
Overlapping users between two domains are defined as $\mathcal{U}^o = \mathcal{U}^s \cap \mathcal{U}^t$. Commonly, $\mathcal{V}^s$ and $\mathcal{V}^t$ do not have overlapping elements in two domains.

The collaborative filtering-based models will transform the users and items into dense embedding vectors for source and target domain respectively. $\mathbf{u}_{i}^s$, $\mathbf{v}_{j}^s$ denote the embedding vectors for $u_i$ and $v_j$ in source domain,  and  $\mathbf{u}_{i}^t$, $\mathbf{v}_{j}^t$ are the corresponding vectors in target domain. The dimensionality of these vectors can be different or the same, here we set they all have $k$  dimensions. The Embedding vectors can be viewed as the outputs of a pre-trained base model with a loss function defined as 
$ \mathcal{L} =  \frac{1}{\vert \mathcal{R} \vert}\sum_{r_{ij} \in \mathcal{R}} \left(  r_{ij} - \mathbf{u}_{i}\mathbf{v}_{j} \right)^2 $.
We note that in addition to predicting ratings, the evaluation metrics of recommender systems can be various such as click-through rate and ranking.

CDRs aim to transfer the knowledge from source domain, and improve the recommendation performance for $\mathcal{U}^t$. In order to evaluate the effectiveness of models, $\mathcal{U}^o$ will be split into a train set and a test set. The ratings of users in train set, which includes both source and target domain interactions, will be used to train a CDR model. For users in the test set, their rating records in target domain will be used for evaluation.

\subsection{Cross-domain Mapping}

For CDR models with embedding-based mapping approach \cite{zhu2021cross}, mapping function is the core module they focus on. We use $f_{map}$ to denote the mapping function. The user embedding in source $ \mathbf{u}_{i}^s$ can be transformed into target domain by $ \hat{\mathbf{u}}_{i}^t = f_{map}(\mathbf{u}_{i}^s) $.
%Mapping optimization, loss for mapping 
To optimize the mapping function, we can define a commonly used loss function to minimize the distance between the transferred user embeddings and the truth user embeddings $ \mathbf{u}_{i}^t$:
\begin{equation}
\mathcal{L}_{map} = \frac{1}{\vert \mathcal{U}^{o}  \vert} \sum_{u_{i} \in \mathcal{U}^{o}} \Vert \hat{\mathbf{u}}_{i}^t - \mathbf{u}_{i}^t \Vert ^{n}
\label{map_loss}
\end{equation}
The parameter $n$ denotes $L_{1}$-norm when $n=1$, or $L_{2}$-norm when $n=2$. It reveals that the motivation of mapping function is having $\hat{\mathbf{u}}_{i}^t$ closer to $ \mathbf{u}_{i}^t$. 

%loss for final task
In terms of the recommendation part, CDR methods will combine $\hat{\mathbf{u}}_{i}^t$ and item embedding in target domain $ \mathbf{v}_{j}^t$ to give a score which indicates the preference of ${u}_{i}$ on ${v}_{j}$.  Here we use the rating score directly and define the task loss function as follows:
\begin{equation}
\mathcal{L}_{task} = \frac{1}{|\mathcal{R}_{o}^{t}|}\sum_{r_{ij} \in  \mathcal{R}_{o}^{t}} \left( r_{ij} -  \hat{\mathbf{u}}_{i}^t \mathbf{v}_{j}  \right)^{2}  
\label{task_loss}
\end{equation}
$\mathcal{R}_{o}^{t} = \{r_{ij}| u_i \in \mathcal{U}^{o}, v_{j} \in \mathcal{V}^{t} \} $ denotes the ratings of overlapping users in target domain.

% Some CDR methods  \cite{man2017cross} only utilize the $\mathcal{L}_{map}$, and ignore the task performance in target domain, while some other models  \cite{zhu2022personalized} only use $\mathcal{L}_{task}$ to optimize the final task.

\subsection{Diffusion Probabilistic Models (DPMs)}
We put the details of DPMs in Appendix (\ref{sec:add}) to keep coherence.

%------------------------------
\section{Model}

\begin{figure}
\centering
\includegraphics[width=0.75\linewidth]{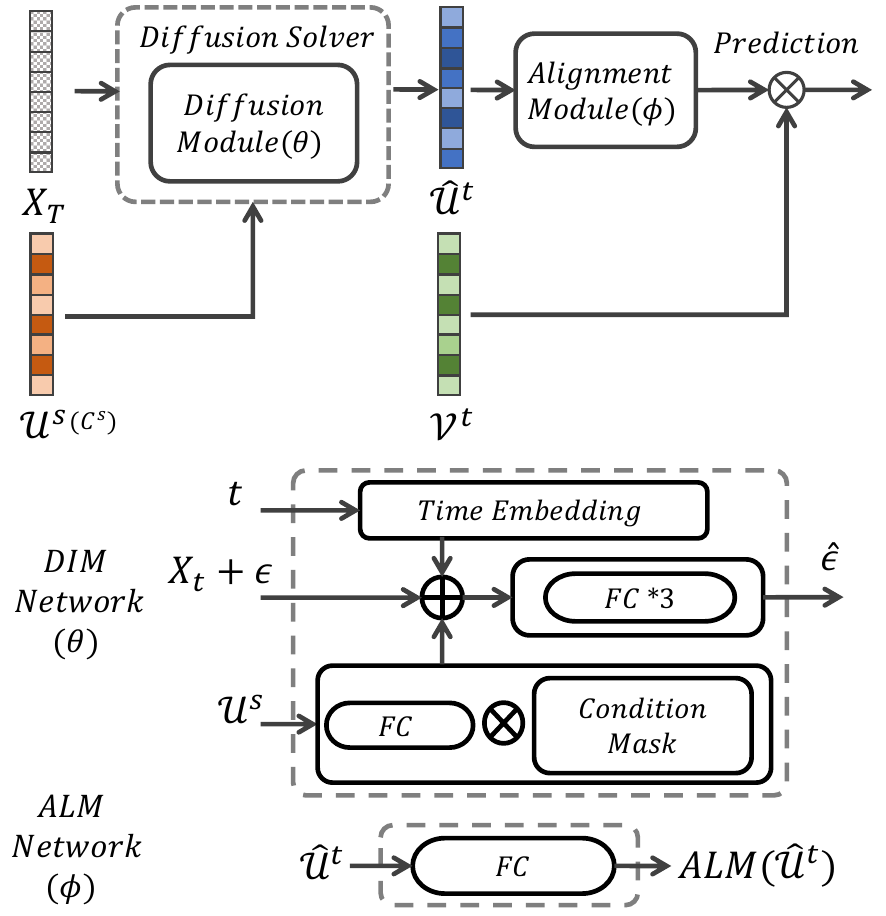}
\caption{Architecture of DiffCDR. It consists of a Diffusion Module (DIM) and an  Alignment Module (ALM). We also employ a diffusion solver to accelerate the inference of the DIM. The recommendation results are produced by element-wisely multiplication with the output of ALM and the item embeddings in target domain. } \label{fig_structure}
\end{figure}

We illustrate the architectures of DiffCDR with the network of DIM and ALM in Figure (\ref{fig_structure}) .

\subsection{Diffusion Module (DIM)}

Starting from random noise $ \mathbf{x}_T \sim \mathcal{N}(0,\mathcal{I})$, we design the DIM to remove the noise via reverse process and produces $ \mathbf{x}_0$ as transferred user embeddings $ \hat{\mathbf{u}}_{i}^t$. It is guided by user's features in source domain. 
For CDR tasks, it is natural to expect $ \hat{\mathbf{u}}_{i}^t$ can  reflect the preference of that user in source domain. 
Based on Equation (\ref{class_free}), we define the classifier-free guidance score function as:
\begin{equation}
\hat{\epsilon}_{\theta}(\mathbf{x}_t| \mathbf{u}_{i}^s ) = \epsilon_{\theta}(\mathbf{x}_t) + s \cdot (\epsilon_{\theta}(\mathbf{x}_t| \mathbf{c}_{i}^{s} ) - \epsilon_{\theta}(\mathbf{x}_t)) 
 \label{class_free_guid_score}
\end{equation}
where $ \mathbf{c}_{i}^{s} $ is the conditional guidance embedding $ \mathcal{C}^s$   of $u_i$ in source domain. 
In each step, we can derive $\mathbf{x_{t-1}}$ from $\hat{\epsilon}_{\theta}(\mathbf{x_t}| \mathbf{u}_{i}^s )$   \cite{ho2020denoising}  and obtain the sample $\mathbf{x}_{t-1}$ by 
$ \mathbf{x}_{t-1} = DIM(\hat{\epsilon}_{\theta}, \mathbf{x}_t , \mathbf{c}_{i}^{s} ,t  )  $.
By utilizing DIM repeatedly, we can have $\mathbf{x}_T,\mathbf{x}_{T-1}, ..., \mathbf{x}_0 $. We use the final step sample $\mathbf{x}_0$ as the transferred user embeddings in target domain, that is  $ \hat{\mathbf{u}}_{i}^t = \mathbf{x}_0 $.

\noindent $\bullet$ \textbf{Structure of DIM.}
%structure is not decided yet
Many image generative diffusion models \cite{li2022diffusion,song2020denoising,kingma2021variational} use U-net \cite{ronneberger2015u} as their backbone. Considering the dimensionality of the data in our tasks, we employ layers of perceptron to construct DIM, which already achieve good performance in our experiments.
%time embedding
In order to be compatible with diffusion solvers, we add sinusoidal position embedding  \cite{vaswani2017attention} to build time step embeddings.
We utilize linear layers to adjust the time step embedding and the conditional guidance to have the same units as the input. We then add these three parts element-wisely to give the estimation of added noise $\hat{\epsilon}$.

\noindent $\bullet$ \textbf{Fast sampling of DIM.}
The inference of DIM can be time-consuming. Instead of a step-by-step evaluation with DIM, we adopt a fast inference solver for diffusion ODEs to speed up the sampling process with much fewer steps:
\begin{equation}
\hat{\mathbf{u}}_{i}^t= \mathbf{x}_{0} = Solver( \hat{\epsilon}_{\theta}, \mathbf{x}_T , \mathbf{c}_{i}^{s}  )
\label{solver_result}
\end{equation}  
We can choose the $Solver$ from existing works  \cite{bao2022analytic,popov2021diffusion,lu2022dpm}.
Moreover, the fast inference of DIM enables us to have a stable training procedure for a specific recommendation task. 

\subsection{Alignment Module (ALM)}
Since the randomness could be introduced by diffusion model both in training and inference phases, we employ a alignment module (ALM) to ensure the consistency between the generated user embedding $\hat{\mathbf{u}}_{i}^t$ from DIM and the ground-truth user representation in target domain $\mathbf{u}_{i}^t$. We then obtain the rating prediction as $ \hat{r}_{ij} =  ALM_{\phi}( \hat{\mathbf{u}}_{i}^t)  \mathbf{v}_{j}^t $. 
The motivation of ALM is to alleviate the effects of randomness, which improves the quality of user preference transferring. 

In our experiments, we implement ALM by simply adding one fully connected layer $\mathrm{FC}(\hat{\mathbf{u}}_{i}^t)$, which takes $\hat{\mathbf{u}}_{i}^t$ as input. We note that the ALM along cannot not give satisfying recommendation results. In the ablation experiments, we analyze the influence of ALM in detail.

\subsection{Loss Computation}
$\bullet$ \textbf{DIM Loss.}
In order to generate a training sample, we choose a $t\in[0,T]$ randomly, and add noise by pre-defined noise schedule on $\mathbf{u}_{i}^t$ as  
\begin{equation}
q(\mathbf{x}_t | \mathbf{u}_{i}^t ) = \mathcal{N}( \mathbf{x}_t ; \bar{\alpha}(t)\mathbf{u}_{i}^t, \sigma^{2}(t)\mathcal{I}) 
\label{diff_train_sample}
\end{equation}

Following the loss defined in (\ref{diff_loss}), the loss function of DIM is:
$$ \mathcal{L}_{DIM_\theta}=\mathcal{L}_{\hat{\epsilon}_{\theta}} = \mathbb{E}_{t \in [0,T],\mathbf{x}_{0} \sim q(  \mathbf{u}_{i}^t  ), \epsilon \sim\mathcal{N}(0,\mathcal{I}) } \left[  || \epsilon - \hat{\epsilon}_{\theta}( \mathbf{x}_t , t | \mathbf{c}_{i}^{s})  || \right],$$
where $\epsilon \sim \mathcal{N}(0,\mathcal{I})$ , and  $\hat{\epsilon}_{\theta} $ is defined in (\ref{class_free_guid_score}). For the purpose of training  $\epsilon_{\theta}( \mathbf{x}_t , t )$ and $\epsilon_{\theta}( \mathbf{x}_t , t | \mathbf{c}_{i}^{s})$ at the same time, we will mask the conditional information $\mathbf{c}_{i}^{s}$ randomly during training phase with a certain ratio.\\

\noindent $\bullet$ \textbf{ALM and Task Loss.}
To form the ALM loss function, we add the ALM into the $\mathcal{L}_{map}$ defined in (\ref{map_loss}):
\begin{equation}
  \mathcal{L}_{ALM_\phi} = \frac{1}{\vert \mathcal{U}^{o}  \vert} \sum_{u_{i} \in \mathcal{U}^{o}}  \Vert ALM_\phi(\hat{\mathbf{u}}_{i}^t) - \mathbf{u}_{i}^t \Vert ^{n}
\label{loss_alm}  
\end{equation}
Following the architecture of PTUPCDR\cite{zhu2022personalized}, we also form a task loss to optimize the our model with ratings data directly. This task-oriented strategy has our model adapt to the specific recommendation tasks. 
Consider both the loss defined in (\ref{task_loss}) , we give the task loss function as 
$$
\mathcal{L}_{task_\phi} = \frac{1}{|\mathcal{R}_{o}^{t}|}\sum_{r_{ij} \in  \mathcal{R}_{o}^{t}}   ( r_{ij} - ALM_\phi(\hat{\mathbf{u}}_{i}^t)\mathbf{v}_{j}  )^{2}.
$$

% Consider both the loss defined in (\ref{task_loss}) and BPR loss\cite{10.5555/1795114.1795167}, we give the task loss function as 
% $$
% \mathcal{L}_{task_\phi} = \frac{1}{|\mathcal{R}_{o}^{t}|}\sum_{r_{ij} \in  \mathcal{R}_{o}^{t}}  \left( \lambda_1 ( r_{ij} - \hat{r}_{ij}  )^{2}
% - \lambda_2 \mathrm{logSigmoid}( \hat{r}_{ij} - neg\_\hat{r}_{ij} ) \right),
% $$
% 
% where $
% \hat{r}_{ij} = ALM_\phi(\hat{\mathbf{u}}_{i}^t)\mathbf{v}_{j},
% $ 
% $
% neg\_\hat{r}_{ij} = ALM_\phi(\hat{\mathbf{u}}_{i}^t)\mathbf{neg\_v}_{j},
% $ and $\mathbf{neg\_v}_{j}$ is a negative sample drawn from train set randomly. $\lambda_1$ and $\lambda_2$ are used to adjust the weights.

Since $\mathcal{L}_{ALM_\phi}$ and  $\mathcal{L}_{task_\phi}$ both depend on the output of DIM and share same trainable variables $\phi$, we optimize these two losses together via:
\begin{equation}
\mathcal{L}_{ALM\_ task_\phi} = \mathcal{L}_{ALM_\phi} +  \lambda\mathcal{L}_{task_\phi} 
\label{loss_total}  
\end{equation}
where $\lambda$ is used to adjust the weight.

% \begin{equation}
% \mathcal{L}_{ALM\_ task_\phi} = \mathcal{L}_{ALM_\phi} +  \mathcal{L}_{task_\phi} 
% \label{loss_total}  
% \end{equation}

We will also analyze the influence of task loss in ablation experiments. 

% 
% Finally, the loss function for DiffCDR is as follows:
% \begin{equation}
% \mathcal{L}_{Total} =
% \mathcal{L}_{DIM_\theta} +  \mathcal{L}_{ALM_{\phi}} + \lambda\mathcal{L}_{task_\phi} ,
% \label{loss_total}
% \end{equation}
% where $\lambda$ are weights to balance the task loss.

\subsection{Training Strategy}
In the training phase, we first need to generate the transferred user embeddings via DIM and then use them as input to ALM. We choose DPM-solver  \cite{lu2022dpmsolver} as the fast inference solver here, which enables us to accelerate the inference process and generate $x_0$ with much fewer steps.  Therefore, we take a training strategy to optimize $\mathcal{L}_{DIM_\theta}$ and $\mathcal{L}_{ALM\_ task_\phi}$ alternately in each batch.

We describe the training procedure formally in Algorithm (\ref{alg1}).

%algorithm process 1
\begin{algorithm}[h] 
\setstretch{0.6}

\caption{Training Procedure of Diffusion Cross-Domain Recommendation (DiffCDR)}   
\label{alg1}  
\begin{algorithmic} [1]

\renewcommand{\algorithmicrequire}{ \textbf{Input:}} 
\REQUIRE $ \mathcal{U}^s,\mathcal{V}^s,\mathcal{R}^s,\mathcal{C}^{s},\mathcal{U}^o,\mathcal{V}^t,\mathcal{R}^t $

\REQUIRE  Time steps $ t \in [0,T]$, Noise schedule $ \bar{\alpha}(t) ,\sigma(t)$.

\REQUIRE Number of training epochs $N$ 
\REQUIRE Diffusion score function $\hat{\epsilon}_{\theta}$, 
\REQUIRE Alignment module $ALM_{\phi} $.

\REQUIRE Fast inference solver $Solver$ .

\renewcommand{\algorithmicrequire}{ \textbf{Pre-training Stage:}} 
\REQUIRE

\STATE{Learning a source model with $\mathbf{u}^{s},\mathbf{v}^{s}$}.
\STATE{Learning a target model with $\mathbf{u}^{t},\mathbf{v}^{t}$}.

%\ENSURE $$   
\renewcommand{\algorithmicrequire}{ \textbf{DiffCDR Training Stage:}} 
\REQUIRE

% \renewcommand{\algorithmicdo}{ \textbf{or converge}} 

%detials:
\STATE  Initializing $\hat{\epsilon}_{\theta}$ randomly
\STATE  Initializing $ALM_{\phi}$ randomly
\FORALL{ $N$ }
    \FORALL{ $Batches$ }

    \STATE 
    Generating training samples by Equation (\ref{diff_train_sample})
      
    \STATE 
    Updating $DIM_\theta$ by minimizing $\mathcal{L}_{DIM_\theta}$

    \STATE 
    Inferring $\hat{\mathbf{u}}^t$ by Equation (\ref{solver_result})

    \STATE 
    Updating $ALM_{\phi} $ by minimizing $\mathcal{L}_{ALM\_ task_\phi}$ 
    \ENDFOR
\ENDFOR

\end{algorithmic}  
\end{algorithm}

%------------------------------
\section{Experiments}
\noindent $\bullet$ \textbf{Research Questions (RQ).}
We intend to answer the following research questions via experiments:
\textbf{RQ1.}How does DiffCDR perform on CDR tasks comparing to state-of-art baseline methods ? 
\textbf{RQ2.}How do DIM and ALM contribute to the whole method? What is the influence of task loss on the performance for DiffCDR ?
\textbf{RQ3.}Why can we have better performance with DiffCDR?
\textbf{RQ4.}As a generative-based method, what is the throughout of DiffCDR ?

\subsection{Experimental Settings}

\begin{table}[h]
    \renewcommand\arraystretch{0.8}
    
  \caption{Data Statistics of CDR tasks. }
  \label{tab_data}
  \resizebox{7 cm}{!}{
  \begin{tabular}{c|p{1.1cm}<{\centering} p{1.1cm}<{\centering}| p{1.1cm}<{\centering} p{1.1cm}<{\centering} |p{1.1cm}<{\centering} p{1.1cm}<{\centering} }
    \toprule
    \multirow{2}{*}{ \makecell{CDR \\ Tasks}}  &  \multicolumn{2}{c|}{Domain} & \multicolumn{2}{c|}{Item} &  \multicolumn{2}{c}{Rating} \\
    
    & Source & Target  & Source & Target &  Source & Target \\
    \midrule
    Task1 & Video & Music & 50 K & 64 K &  1,697 K & 1,097 K \\
    
    Task2 & Book & Video & 368  K & 50  K  & 8,898 K & 1,698  K \\

    Task3 & Book & Music & 368  K & 64 K  & 8,898 K &  1,097 K \\
    \bottomrule
  \end{tabular}
  }
  
  \resizebox{7 cm}{!}{
  \begin{tabular}{c|p{1.1cm}<{\centering} p{1.1cm}<{\centering} |p{1.6cm}<{\centering} p{1.6cm}<{\centering} p{1.5cm}<{\centering}}
    \toprule
    \multirow{2}{*}{\makecell{CDR \\ Tasks}}  &  \multicolumn{2}{c|}{Domain}  & \multicolumn{3}{c}{User}  \\
    
    & Source & Target  &  Overlap  & Source & Target \\
    \midrule
    Task1 & Video & Music &  18 K & 124 K & 75 K  \\
    
    Task2 & Book & Video &  37 K &  604 K &  124 K   \\
    
    Task3 & Book & Music &  17 K &  604 K &  75 K  \\
    \bottomrule
  \end{tabular}
  }
\end{table}
\noindent $\bullet$ \textbf{Dataset.}
Following the previous works \cite{kang2019semi,zhao2020catn,zhu2022personalized,10.1145/3459637.3482137}, we utilize three real-word public datasets built on Amazon review data\footnote{http://jmcauley.ucsd.edu/data/amazon/}. We use the 5-core subsets, in which all users and items have at least 5 reviews. And we choose 3 categories with large data volume from all 24 ones, which are movies\_and\_tv (Video), cds\_and\_vinyl (Music), and books (Book). 
We set three CDR tasks on Amazon dataset: Task 1: Video → Music, Task 2:
Book → Video, and Task 3: Book → Music.  The data statistics are in Table (\ref{tab_data}).

\noindent $\bullet$ \textbf{Metrics.} 
Since the Amazon reviews data contains the rating records (0-5), it is a typical regression problem. We follow  \cite{kang2019semi,zhao2020catn,zhu2022personalized} and adopt the metrics including  Mean Absolute Error
(MAE) and Root Mean Square Error (RMSE). In addition, we report the Normalized DCG (nDCG) and hit rate for highest 20 scores(N@20, H@20). Noting that the nDCG and hit rate scores are calculated by ranking in all target domain items. To obtain these ranking metrics, we conduct additional experiments by adding same negative samples to each CDR models.
For each experiment, we report the mean values of five runs.

\noindent $\bullet$ \textbf{Construction Datasets.}
In terms of data split, we follow a similar strategy as  \cite{zhu2022personalized}.
First, we randomly select a fraction of overlapping users as test users, and the remaining overlapping users as training users. We utilize their rating records to train CDR baseline models and our DiffCDR. 
We set the proportion of test users $\beta$ as three cold-start levels: 20\%, 50\% and 80\% of all overlapping users. The larger the $\beta$ is, the more difficult a CDR task will be.
In the cold-start experiments, the ratings of test users are not available during training. 
For the  warm-start experiments, we divide 50\% of test users' rating records as fine-tune data. During that data split operation, we consider the order of time series to avoid potential leakage.

\noindent $\bullet$  \textbf{Baselines.}
 \textbf{(1) TGT} denotes a simple MF model, which is only trained on target domain data.
\textbf{(2) CMF}  \cite{singh2008relational} shares same embeddings for overlapping users belonging to different domains. It is collectively trained on both datasets from source and target domains.
 \textbf{(3) EMCDR}  \cite{man2017cross} is a classic CDR model which adopts a network as the mapping function to transfer the user embeddings from source domain into target domain. The initial user embeddings in this method are produced by two separate MF models that are trained in two domains separately.
 \textbf{(4) SSCDR} \cite{kang2019semi} is an embedding-based mapping approach model which leverages semi-supervised method.  
\textbf{(5) LACDR} \cite{10.1145/3459637.3482137} employs encoder-decoder structure to form the mapping function. It aligns the low-dimension embedding spaces of different domains.
 \textbf{(6) PTUPCDR} \cite{zhu2022personalized} is a meta mapping approach CDR method. It generates personalized mapping functions (bridge functions) by a meta-network for each user, which transfer knowledge across domains. We discuss VAE framework CDRs in Section \ref{sec:vae} and explain why we do not include VAE-based baselines.

\noindent $\bullet$ \textbf{Experimental Steps.}
We conduct our experiments in the following steps:  \textbf{(1)} Train all models on the training dataset which only includes training users. In each group of experiments, we use the same pre-trained base models to produce embedding variables.  \textbf{(2)} Evaluate the exact cold-start performance of models on test datasets.   \textbf{(3)} Finally, we will fine-tune the models with warm-start samples, and evaluate the warm-start performance.

\noindent $\bullet$ \textbf{Hyper-parameters Setting.}
%1. embedding related
%2.training sampling related
%3. taining: loss function related 
%4.inference
%5.
\textbf{(1)} We set the dimensionality to 10 for both user and item embedding variables.
\textbf{(2)} We use the Adam optimizer and set learning rate as 0.001.
\textbf{(3)} To train the  score function with and without guidance simultaneously, the overall masking probability for condition input is 10\%.
\textbf{(4)} We set the weight parameter $\lambda $ in Equation (\ref{loss_total})  by grid search among [0.01,0.1,1,2,4,8]. The parameter $n$ in the Equation (\ref{loss_alm}) is set as $n=1$, which obtains better performance than $n=2$.
% \textbf{(4)} We set the weight parameter $\lambda_1, \lambda_2$ in $ \mathcal{L}_{task_\phi}$ by grid search among [0.01,0.1,1,2,4,8]. The parameter $n$ in the Equation (\ref{loss_alm}) is set as $n=1$, which obtains better performance than $n=2$.
\textbf{(5)} Instead of Gaussian noise, we directly utilize source-domain user embeddings as input to DIM for inference phase in our experiments. We find that it can further reduce the number of inference steps while maintaining similar performance. 
\textbf{(6)} In terms of the architecture of DIM, we use 3 layers of perceptrons with 128 hidden units ( 10 units for input and output ). For the ALM, we utilize one fully connected layer without bias variable, which has the same input and output dimensionality as user embedding(=10) .
\textbf{(7)} The value of guidance strength $s$ in Equation (\ref{class_free_guid_score}) is set as 0.1 in our experiments, which is decided by grid search from [0.05,0.1,0.2,0.5,1,2,5]. The number of function evaluations used in DPM-solver is 30, which trades off quality against inference speed.

\subsection{Cold- and Warm-start Experiments (RQ1)}

We present the results of cold- and warm-start experiments and analyze the performance of DiffCDR for each scenario in this section.

\begin{table*}[ht]
   \renewcommand\arraystretch{0.7}

  \caption{Metrics of cold-start experiments. Each result is an average of five runs. * means significant difference by paired t-test with $\rho = 0.05$ compared with the best baseline. The column of "Improve" indicates the rate of improvement. }
  
  \label{tab_cold-star}
  \resizebox{14 cm}{!}{
  \begin{tabular}{c|c|c| p{1.4cm}<{\centering} p{1.6cm}<{\centering} p{1.4cm}<{\centering} p{1.4cm}<{\centering} p{1.4cm}<{\centering} p{1.4cm}<{\centering} p{1.4cm}<{\centering} | p{1.4cm}<{\centering}}
    \toprule
      & $\beta$ & Metrics & TGT & CMF & EMCDR & SSCDR &  LACDR & PTUPCDR & DiffCDR & Improve \\
    \midrule
    \multirow{12}{*}{ \makecell{Task1 \\  Video \\ ->\\  Music} } & \multirow{4}{*}{20\%} & MAE & 4.4546 & 1.4642 & 1.3596 & 1.1757 & 1.1295 & 1.1099 & \textbf{1.0435*} & \textbf{6.0\%} \\
                                                 & & RMSE& 5.1338 & 1.9571 & 1.6615 & 1.4911 & 1.4358 & 1.4543 & \textbf{1.3840*}  & \textbf{3.6\%}\\ 
                                                 & & N@20 & 0.00253 & 0.00508 & 0.00977 & 0.00932 & 0.00984 & 0.00978 & \textbf{0.01026*} & \textbf{4.3}\%\\
                                                 & & H@20 & 0.00033 & 0.00084 & 0.00229 & 0.00212 & 0.00228 & 0.00236 & \textbf{0.00238*} & \textbf{1.1}\%\\
                                                \cline{2-11}
                           & \multirow{4}{*}{50\%} & MAE & 4.4884 & 1.6710 & 1.6891 & 1.4320 & 1.3502 & 1.2842 & \textbf{1.2367*} & \textbf{3.7\%}\\
                                                 & & RMSE& 5.1790 & 2.2076 & 2.0368 & 1.8248 & 1.7510 & 1.7340 & \textbf{1.6859*}  & \textbf{2.8\%}\\
                                                 & & N@20 & 0.00251 & 0.00403 & 0.00898 & 0.00793 & 0.00893 & 0.00828 & \textbf{0.00915*} & \textbf{1.9\%}\\
                                                 & & H@20 & 0.00033 & 0.00068 & 0.00193 & 0.00164 & 0.00199 & 0.00179 & \textbf{0.00202*} & \textbf{1.9\%}\\
                                                 \cline{2-11}
                           & \multirow{4}{*}{80\%} & MAE & 4.4959 & 2.2327 & 2.1980 & 1.8162 & 1.6886 & 1.6174 & \textbf{1.5606*}  & \textbf{3.5\%}\\
                                                 & & RMSE& 5.1830 & 2.8868 & 2.5713 & 2.3090 & 2.2238 & 2.2429 & \textbf{2.1754*}  & \textbf{2.2\%}\\
                                                 & & N@20 & 0.00248 & 0.00348 & 0.00622 & 0.00578 & 0.00606 & 0.00545 & \textbf{0.00665*} & \textbf{6.9\%}\\
                                                 & & H@20 & 0.00033 & 0.00051 & 0.00124 & 0.00111 & 0.00124 & 0.00107 & \textbf{0.00136*} & \textbf{9.7\%} \\
    \hline
    \multirow{12}{*}{\makecell{Task2 \\  Book \\ ->\\  Video} } & \multirow{4}{*}{20\%} & MAE & 4.1807 & 1.4742 & 1.1305 & 0.9774 & 0.9681 & 1.0728 & \textbf{0.9476*} & \textbf{2.1\%} \\
                                                 & & RMSE& 4.7496 & 1.9180 & 1.4215 & 1.2356 & \textbf{1.2311} & 1.3745 & 1.2338  & -0.2\% \\
                                                 & & N@20 & 0.00245 & 0.00578 & 0.01898 & 0.02066 & 0.01850 & 0.01821 & \textbf{0.02073} & \textbf{0.3\%}\\
                                                 & & H@20 & 0.00043 & 0.00124 & 0.0064 & 0.00676 & 0.0056 & 0.00594 & \textbf{0.00697*} & \textbf{3.1\%} \\
                                                \cline{2-11}
                           & \multirow{4}{*}{50\%} & MAE & 4.1951 & 1.5651 & 1.1863 & 1.0193 & 1.0077 & 1.1116 & \textbf{0.9953} & \textbf{1.2\%} \\
                                                 & & RMSE& 4.7693 & 2.0341 & 1.4993 & 1.3089 & \textbf{1.3051} & 1.4425 & 1.3155 & -0.8\% \\
                                                 & & N@20 & 0.00274 & 0.00536 & 0.01924 & 0.02041 & 0.01875 & 0.01785 & \textbf{0.02047} & \textbf{0.3\%} \\
                                                 & & H@20 & 0.00044 & 0.00107 & 0.00642 & 0.00675 & 0.00535 & 0.00575 & \textbf{0.0068} & \textbf{0.7\%} \\
                                                 \cline{2-11}
                           & \multirow{4}{*}{80\%} & MAE & 4.2384 & 2.2379 & 1.3445 & 1.1469 & 1.1151 & 1.2072 & \textbf{1.0846*} & \textbf{2.7\%} \\
                                                 & & RMSE& 4.8198 & 2.7538 & 1.6946 & 1.4871 & \textbf{1.4660} & 1.5987 & 1.4695 & -0.2\% \\
                                                 & & N@20 & 0.00258 & 0.00412 & 0.01906 & 0.01949 & 0.01710 & 0.01520 & \textbf{0.01960} & \textbf{0.6\%} \\
                                                 & & H@20 & 0.00040 & 0.00073 & 0.00628 & \textbf{0.00636} & 0.00512 & 0.00484 & 0.00634 & -0.3\% \\
    \hline       
    \multirow{12}{*}{\makecell{Task3 \\ Book \\ -> \\ Music}} & \multirow{4}{*}{20\%} & MAE & 4.5190 & 1.7976 & 1.6425 & 1.3073 & 1.1945 & 1.2556 & \textbf{1.1220*} & \textbf{6.1\%} \\
                                                 & & RMSE& 5.1838 & 2.3545 & 1.9873 & 1.6599 & 1.5771 & 1.6730 & \textbf{1.5390*} & \textbf{2.4\%} \\
                                                 & & N@20 & 0.00196 & 0.00383 & 0.01193 & 0.01179 & 0.01367 & 0.01006 & \textbf{0.01374} & \textbf{0.5\%} \\
                                                 & & H@20 & 0.00035 & 0.00071 & 0.00323 & 0.00313 & 0.0037 & 0.00275 & \textbf{0.00382*} & \textbf{3.2\%} \\
                                                \cline{2-11}
                           & \multirow{4}{*}{50\%} & MAE & 4.4953 & 2.0002 & 1.9364 & 1.5183 & 1.3925 & 1.4304 & \textbf{1.3077*} & \textbf{6.1\% }\\
                                                 & & RMSE& 5.1685 & 2.6001 & 2.2966 & 1.9467 & 1.8644 & 1.9475 & \textbf{1.8255*} & \textbf{2.1\%} \\
                                                 & & N@20 & 0.00200 & 0.00341 & 0.00994 & 0.00964 & 0.01058 & 0.00804 & \textbf{0.01082*} & \textbf{2.3\%} \\
                                                 & & H@20 & 0.00028 & 0.00059 & 0.00253 & 0.00247 & 0.00277 & 0.00206 & \textbf{0.00281*} & \textbf{1.7\%} \\
                                                 \cline{2-11}
                           & \multirow{4}{*}{80\%} & MAE & 4.5133 & 2.5014 & 2.3448 & 1.8849 & 1.7107 & 1.7016 & \textbf{1.5871*} & \textbf{6.7\%} \\
                                                 & & RMSE& 5.1960 & 3.1740 & 2.7035 & 2.3517 & 2.2468 & 2.3248 & \textbf{2.2110*} & \textbf{1.6\%} \\
                                                 & & N@20 & 0.00170 & 0.00275 & 0.00705 & 0.00652 & 0.00658 & 0.00682 & \textbf{0.00722*} & \textbf{2.3\%} \\
                                                 & & H@20 & 0.00027 & 0.00046 & 0.00176 & 0.00158 & 0.00165 & 0.00107 & \textbf{0.00179*} & \textbf{1.6\%} \\
    \bottomrule
  \end{tabular}
  }
\end{table*}

\noindent $\bullet$ \textbf{Cold-start Experiments.} Following the prior works \cite{zhu2022personalized}, we evaluate DiffCDR in 3 CDR tasks with different values of $\beta$. We have several insights from the experimental results in Table (\ref{tab_cold-star}).
\textbf{(1)} TGT only takes the data from target domain, that means the embedding variables of test users are untrained due to data sparsity. As a result, TGT falls behind in the experiments. In contrast, other methods utilize data from both domains and achieve better results. Consequently, exploiting additional data from auxiliary domain is an effective way to promote the recommendation quality.
\textbf{(2)} CMF is a variant of TGT, it combines the data from source and target domains. Thus its embedding space is shared across domains, and it obtains better results than TGT model. Due to the differences in the embedding spaces between two domains and CMF does not consider this fact, CDRs outperform the CMF in our experiments. For CDRs, they transfer the knowledge from source to target domain through various mapping functions and  alleviate the discrepancy of embedding spaces.
It is valuable to design a certain mechanism to extract supplementary information from source domain to boost the performance of cold-start recommendation.
\textbf{(3)} Task2 may be easier than the other two, such that the differences between models are close, but DiffCDR still achieves comparable performance on it. All results show the advantage of DiffCDR in cold-start CDR tasks.

\begin{figure*}
\centering
\includegraphics[width=0.75 \linewidth]{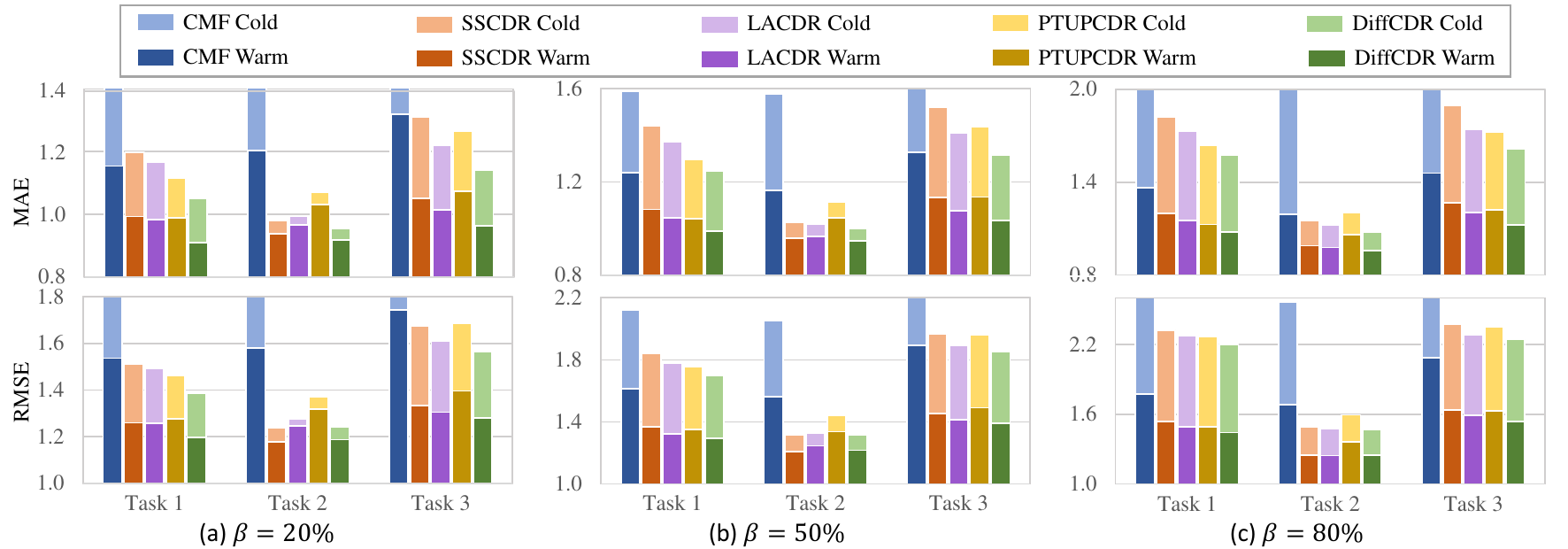}
\caption{Results of warm-start experiments. The light and dark colored represent the cold-start and warm-start respectively. } \label{fig_warm-start}
\end{figure*}

\noindent $\bullet$ \textbf{Warm-start Experiments.} The warm-start scenario represents a situation that a fraction of the interactions from test users are available during the training phase, which we can expect to encounter frequently in real-world applications. It can be viewed as a pre-trained stage for initialization  \cite{zhu2022personalized}. With the CDR methods, this initialization can help a recommender system to generate better results. 

We validate CMF,SSCDR, LACDR, PTUPCDR and our DiffCDR. From the results in Figure \ref{fig_warm-start}, we have insights as follows: 
\textbf{(1)} Additional data can be helpful to improve recommendation performance. All models have better metrics than the cold-start scenario. 
\textbf{(2)} CDRs not only achieve better performance for cold-start tasks, they also can improve the results further with more interaction data . The results show that CDRs are valuable for real-world applications 
\textbf{(3)} Again, it is difficult to gain further improvement on task 2, but DiffCDR still has better results in all experiments. Together with Table (\ref{tab_cold-star}), they demonstrate the effectiveness and adaptability of DiffCDR in both cold-start and warm-start scenarios.

\subsection{Ablation Experiments (RQ2)}

% Ablation1: ALM + TASK loss 
% Ablation2: DIM + ALM 
% Full: DIM + ALM + TASK loss 

\begin{table*}[ht]
\renewcommand\arraystretch{0.9}

  \caption{ Metrics of ablation experiments. $^\dag$ "Best CDR M" denotes the best results of CDR models without target label learning, including CMF,EMCDR,SSCDR and LACDR. $^\ddag$ "DA" denotes the DiffCDR only includes DIM and ALM, "AT" denotes the DiffCDR only includes ALM and target label data learning, while "DAT" denotes the full version of DiffCDR.}

  \resizebox{17 cm}{!}{
  \begin{tabular}{c|c| p{0.75cm}<{\centering} p{0.75cm}<{\centering} p{0.75cm}<{\centering} | p{0.75cm}<{\centering} p{0.75cm}<{\centering} p{0.75cm}<{\centering} | p{0.75cm}<{\centering} p{0.75cm}<{\centering}  p{0.75cm}<{\centering} || p{0.75cm}<{\centering} p{0.75cm}<{\centering} p{0.75cm}<{\centering} | p{0.75cm}<{\centering} p{0.75cm}<{\centering} p{0.75cm}<{\centering} | p{0.75cm}<{\centering} p{0.75cm}<{\centering}  p{0.75cm}<{\centering}}
    \toprule
    \multicolumn{2}{c|}{Metric} & \multicolumn{9}{c||}{ MAE } & \multicolumn{9}{c}{ RMSE } \\
    \midrule
    \multicolumn{2}{c|}{ $\beta$ } & \multicolumn{3}{c|}{ 20\% } & \multicolumn{3}{c|}{ 50\% } & \multicolumn{3}{c||}{ 80\% } & \multicolumn{3}{c|}{ 20\% } & \multicolumn{3}{c|}{ 50\% } & \multicolumn{3}{c}{ 80\% } \\
    \midrule
    \multicolumn{2}{c|}{CDR Task} &  Task1  & Task2 & Task3 &  Task1  & Task2 & Task3 &  Task1  & Task2 & Task3 &  Task1  & Task2 & Task3 &  Task1  & Task2 & Task3 &  Task1  & Task2 & Task3 \\
    \midrule
    \multirow{2}{*}{ \makecell{ Mapping- \\ Oriented} } & Best CDR M$^\dag$ & 1.1295 & 0.9681 & 1.1945 & 1.3502 & 1.0077 & 1.3925 & 1.6886 & 1.1151 & 1.7107 & 1.4358 & \underline{\textbf{1.2311}} & \underline{1.5771} & 1.7510 & \underline{\textbf{1.3051}} & \underline{1.8644} & \underline{2.2238} & \underline{\textbf{1.4660}} & 2.2468 \\
                                                       & DiffCDR DA$^\ddag$ & \underline{1.0843} & \underline{0.9506} & \underline{1.1843} & \underline{1.2764} & \underline{0.9976} & \underline{1.3549} & \underline{1.6458} & \underline{1.1007} & \underline{1.6425} & \underline{1.4157} & 1.2354 & 1.5961 & \underline{1.7241} & 1.3175 & 1.8650 & 2.2259 & 1.4770 & \underline{2.2426}\\
    \hline
    \multirow{2}{*}{ \makecell{ Task- \\ Oriented} } & PTUPCDR & \underline{1.1099} & \underline{1.0728} & \underline{1.2556} & \underline{1.2842} & \underline{1.1116} & \underline{1.4304} & 1.6174 & \underline{1.2072} & 1.7016 & \underline{1.4543} & \underline{1.3745} & \underline{1.6730} & 1.7340 & \underline{1.4425} & 1.9475 & 2.2429 & 1.5987 & 2.3248\\
                                                       & DiffCDR AT & 1.1184 & 1.0912 & 1.2790 & 1.2888 & 1.1257 & 1.4311 & \underline{1.6002} & 1.2093 & \underline{1.6814}  & 1.4597 & 1.3947 & 1.6995 & \underline{1.7321} & 1.4581 & \underline{1.9377} & \underline{2.2019} & \underline{1.5968} & \underline{2.2822} \\
    \hline
    Mapping \& Task & DiffCDR DAT& \textbf{1.0435} & \textbf{0.9476} & \textbf{1.1220} & \textbf{1.2367} & \textbf{0.9953} & \textbf{1.3077} & \textbf{1.5606} & \textbf{1.0846} & \textbf{1.5871} & \textbf{1.3840} & 1.2338 & \textbf{1.5390} & \textbf{1.6859} & 1.3155 & \textbf{1.8255} & \textbf{2.1754} & 1.4695 & \textbf{2.2110}\\

    \bottomrule
  \end{tabular}
  }
\label{tab_ablation}
\end{table*}

To show the contributions of DIM, ALM and task loss, we conduct ablation experiments of cold-start scenario. 
\textbf{Ablation 1: DiffCDR DA.} We removes task loss from DiffCDR and only remain DIM and ALM ("DA") to have fair comparison to the mapping-oriented baselines (denoted as "CDR M", including CMF,EMCDR,SSCDR and LACDR) which are not directly optimized on target domain label data. Here $n$ in the Equation (\ref{loss_alm}) is set as $n=2$ which obtains better results.
\textbf{Ablation 2: DiffCDR AT.} We only includes ALM and task loss learning ("AT") in DiffCDR to illustrate the effect of DIM. 
Since a single ALM is equivalent to EMCDR, we do not conduct separate ablation experiments for this setting.
Noting that we do not conduct additional experiments on ALM module, since the ALM alone is equivalent to EMCDR\cite{man2017cross}.

We summarize the observation of results from Table (\ref{tab_ablation}) as follows. \textbf{(1)} Since label data could include noises, target label data learning methods do not always have better performance. The effect of task-orientated methods varies according to different setting of $\beta$ and tasks. \textbf{(2)} Even without DIM, DiffCDR DA already achieves at least comparable performance to other mapping-oriented CDRs. \textbf{(3)} DiffCDR DAT includes all modules which both consider the mapping and task factors. It further improves the performance of DiffCDR DA. Based on these ablation experiments, DIM, ALM and task label learning are all critical to contribute to the superior performance of DiffCDR.

\subsection{Improvement Analysis (RQ3)}

\begin{figure}
\centering
\includegraphics[width=0.8 \linewidth]{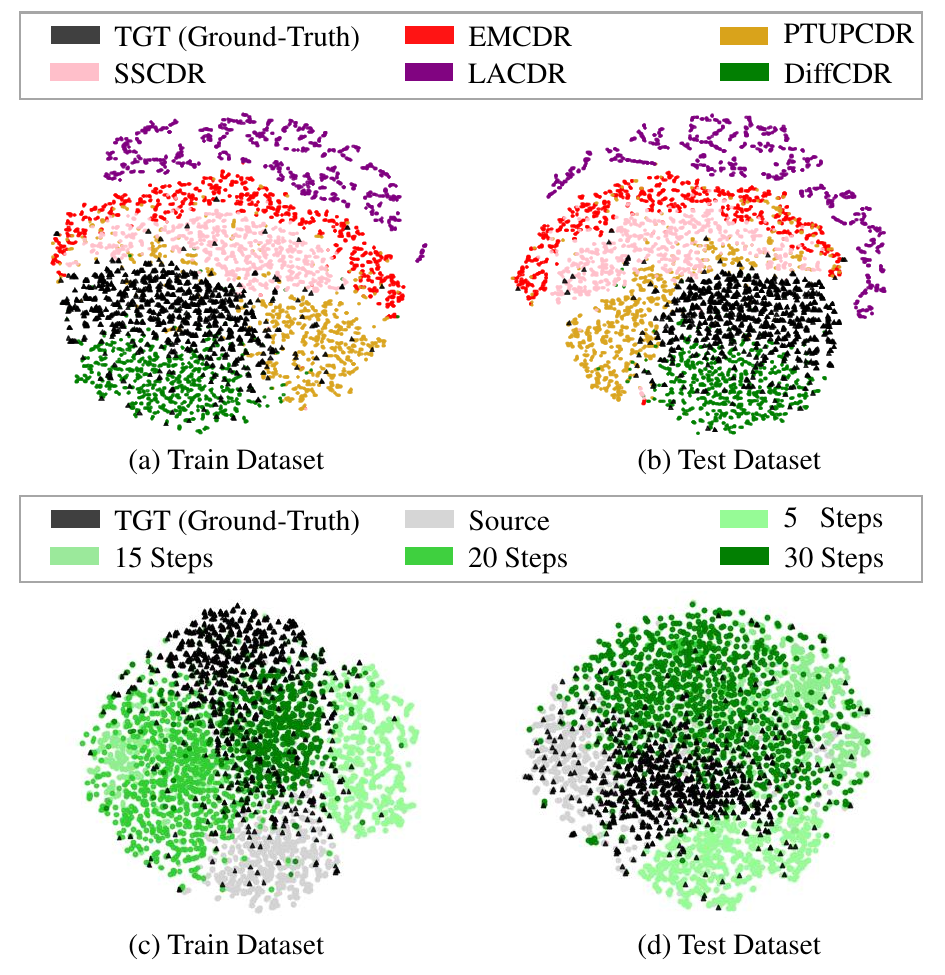}
\caption{User embedding visualization. We randomly sample 1000 users and visualize embeddings of these users in both train and test stages. (a,b) involve different CDRs, and (c,d) are produced by DiffCDR with various inference steps.  } \label{fig_visualized_emb}
\end{figure}

In this section, we analyze the user embeddings generated by four CDRs including our DiffCDR. We reveal the relation between the ground-truth embeddings in target domain and transferred factors.

We use t-SNE \cite{van2008visualizing} implemented in Scikit-learn package to visualize user embeddings in Task1 with $\beta=0.2$. 
To obtain the best user latent factors as ground-truth, we add the test users into train set and train an additional TGT model. The setting of data splitting keeps unchanged for other CDR models. 
For DiffCDR, we derive the user embeddings directly from DIM rather than the output of ALM, since we prefer to show the ability of knowledge transfer across domains. 
We randomly sample 1000 users and all CDRs share the same base models.

\noindent $\bullet$ \textbf{Comparison of CDRs. }
Figure \ref{fig_visualized_emb}(a,b) visualize the embeddings of users in training and test set respectively. Each color denotes a CDR model, and black triangles are produced by TGT. Our observations are as follows.
\textbf{(1)} A satisfied CDR model is able to transfer the factors from one  domain to another, and the transferred factors can be fused with the target data points.
 Although the boundaries between transferred user features and the factors generated by TGT in figures are obvious, DiffCDR (green) has a better fusion with ground-truth (black).
 For other baselines, either the transferred user features are clearly separated from black triangles or there is a large distance.
We can see similar patterns in the train Figure \ref{fig_visualized_emb}(a) and test dataset Figure \ref{fig_visualized_emb}(b). 
\textbf{(2)} The group distance between TGT and DiffCDR is the shortest compared to other methods, it implies that our DiffCDR has the best result in transferring the knowledge into target domain. The most important reason why DiffCDR achieves competitive results is that the distribution of transferred user embeddings by DiffCDR is closer to the latent factors of target domain. 

\noindent $\bullet$ \textbf{Comparison of  Inference Steps with DiffCDR. } To illustrate the effects of inference step, we visualize the user embedding from DiffCDR with various setting of function evaluation numbers in DPM-solver as Figure \ref{fig_visualized_emb} (c,d). The grey dots are samples from source domain. They show that the distance between the transferred and ground-truth user embeddings shrinks as the number of inference steps increases. It also demonstrates how DIM transfers user embeddings across domains during the reverse diffusion process.

\subsection{Throughput Analysis (RQ4)}
\begin{table}[h]
\renewcommand\arraystretch{0.8}
  \caption{ Throughput Statistics of CDR (K samples/second) }
  \label{tab_time_con}
  \resizebox{7 cm}{!}{
  \begin{tabular}{c|p{1.1cm}<{\centering} p{1.1cm}<{\centering} p{1.1cm}<{\centering} p{1.1cm}<{\centering} p{1.1cm}<{\centering} }
    \toprule
      Phrase   & EMCDR & SSCDR  & LACDR & PTUPCDR &  DiffCDR \\
    \midrule
    Training & 39.2 K & 8.3 K & 7.5 K & 32.7 K & 2.9 K \\
    Test  & 125.1 K & 133.4 K & 131.5 K & 100.6 K & 61.5 K\\
    \bottomrule
  \end{tabular}
  }
\end{table}
We report the mean values of CDRs throughput in Table (\ref{tab_time_con}). As generative model, DPM involves time consuming sampling process. With fast diffusion inference solver, our DiffCDR has acceptable throughput. The statistics show that DiffCDR has 38.7\% throughput of SSCDR for training and 61.1\% throughput of PTUPCDR for inference. It is worth noting that we conduct our experiments with 1 core of 2GHz Xeon(R) CPU and 1 core of P100 GPU. Therefore, we could reduce the time consumption further by sampling in parallel.

%------------------------------
\section{Related Work}
\subsection{Mapping Approach CDRs}

Mapping approach CDRs  \cite{zhu2021cross} mainly transfer pre-trained embeddings across domains by various methods. 
EMCDR  \cite{man2017cross} learns a mapping function to infer latent factors for new items/users in target domain. 
DCDCSR \cite{zhu2020deep} considers the sparsity degree to construct benchmark factors with top-K similar entities first, and trained a model to map latent factors into these factors. 
SSCDR  \cite{kang2019semi} is a semi-supervised method, it consists of a supervised learning module trained on overlapping user interactions in different domains, and an unsupervised module that learns from records of user and item interactions in the source domain.
LACDR \cite{10.1145/3459637.3482137} employs encoder-decoder architecture, which includes separate encoders and decoders for different domains. It minimizes the reconstruction errors for each domain and the alignment errors of the low-dimensional hidden variables produced by encoders across domains. 

%meta learning

Inspired by the successful applications of meta-learning, researchers have proposed a number of meta-based CDRs recently. 
MWUF  \cite{zhu2021learning} takes item features and interacted users' features to generate a meta scaling network and a meta shifting network respectively, which warms up cold-start items' embedding.
TMCDR  \cite{zhu2021transfer}  adopts meta learning to alleviate over-fitting problems on limited data of overlapping users. It divides the samples of overlapping users into two groups to construct training tasks, which contain a learning phase and a cold-start phase. 
PTUPCDR  \cite{zhu2022personalized} is a method which considers the various preferences for each user. Instead of learning a shared constant mapping function, this model trains a meta network which generates a personalized mapping function (bridges) for each user.

Our DiffCDR belongs to the class of mapping approach CDRs as well. We leverage probabilistic diffusion model as a powerful method for distribution transforming. It transfers knowledge by generating user embeddings in target domain conditioned on source domain guidance.

%other 

In addition, there are also many other works to solve CDR falling into mapping approach, but we cannot compare with them directly.  
CATN \cite{zhao2020catn} and RC-DFM\cite{10.1609/aaai.v33i01.330194} leverage reviews of users to  extract aspect level information, and performs recommendation by aspect correlations across domains. 
DACIR  \cite{wu2022dynamics} aligns the cross-domain embeddings with a reinforcement learning based interactive recommender method, which takes both the static and dynamic interaction patterns into consideration.
One-to-many transfer method  \cite{krishnan2020transfer} adopts meta transferring to implement CDRs on multiple domains rather than apply separate models.
Ma et al. \cite{10.1145/3331184.3331200} propose a $\pi-$Net which contains a shared account filter unit and a cross-domain transfer unit. This model gives recommendations sequentially for same users in different domains.

\raggedbottom

\subsection{VAE Framework CDRs}
\label{sec:vae}
Recently, variational autoencoders (VAEs) based recommendation methods have achieved excellent results   \cite{liang2018variational, shenbin2020recvae,truong2021bilateral}.
In terms of CDR problems, VAE framework-based models focus on the alignment of latent embedding spaces across domains, which improves the recommendation quality by exploiting domain-sharing information.
SA-VAE \cite{salah2021towards} utilizes a trained VAE model in source domain to operate soft-alignment, and ensures the variational models in different domains share similar latent spaces.
CDRIB  \cite{cao2022cross} creates global graph structure of users and items by variational graph encoder and creates domain-shared representations with information bottleneck regularizers.
VDEA \cite{liu2022exploiting} extracts user preference by variational inference first, and then generates user groups to cluster users with similar rating behaviors. Both information captured from overlapping users and non-overlapping users can be exploited to improve cross-domain performance in this method.

The VAE framework CDRs need to simultaneously approximate the posterior distribution and optimize recommendation task. For that reason, VAE framework CDRs cannot directly utilize user representations from pre-trained models. Therefore, it is difficult to have a controlled comparison with our method, even if both VAEs and DPMs are generative models.

\subsection{Diffusion Probabilistic Models (DPMs)}
To train a generative model, VAEs normally need to build a posterior model at the same time. GANs require an additional discriminator network in addition to the generator, and Normalizing Flow models often require invertible transformation functions. In contrast to those methods, DPMs only need to train a generator to fit each step of reverse diffusion process. As a result, DPMs have fewer constraints on the choice of model architecture.

Following the early work \cite{sohl2015deep} of this topic, DDPM  \cite{ho2020denoising} achieves impressive success in image generation tasks. After that,conditional DPM \cite{dhariwal2021diffusion} improves the performance of image synthesis tasks by class-conditional generation. Classifier-free guided DPM \cite{ho2022classifier} provides generation guidance without a separate classifier model. Large scale image generation APIs based on DPMs also have become popular \cite{ramesh2021zero,nichol2021glide,rombach2022high} recently.
Besides image synthesis, researchers have proposed innovative diffusion models in other fields \cite{kong2020diffwave,han2022card,li2022diffusion}.

DiffRec\cite{wang2023diffusion} leverages diffusion model recommendation tasks. It corrupts  users’ historical interactions in forward steps and recover original interactions iteratively in reverse process. The recovered results are used to recommend non-interacted items.  Since we focuses on the CDR problem, the application and motivation of DiffRec are different from our approach.

%accelerate
%need more

Time-consuming inference is one obvious deficiency of DPMs. In order to deal with this it, DDIM \cite{song2020denoising} adopts non-Markovian noising process to reduce the steps in inference process.
Distillation is also an effective solution to accelerate the sampling procedure \cite{salimans2022progressive}.
When we consider the DPMs from continues-time aspect \cite{song2020score}, researchers propose SED solvers \cite{bao2022analytic,popov2021diffusion} and  ODE solvers  \cite{lu2022dpm,lu2022dpmsolver} for DPMs, which can be used to perform inference dozens of times faster.

% In this paper, we mainly leverage the technique of classifier-free guided DPM \cite{ho2022classifier} with DPM-Solver \cite{lu2022dpm}, thus we transfer the knowledge across domains with high efficiency, and improve CDR performance significantly.

%------------------------------
\section{Conclusion and Future Work}
We study the purpose and existing methods of CDR in this paper first, and explain the mechanism of CDRs with mapping approach in detail. we also investigate DPMs in image synthesis applications, and reveal the motivations of combining DPMs with recommendation models to solve CDR tasks. We then propose a novel Diffusion Cross-domain CDR (DiffCDR) method. In our DiffCDR, a DIM is in charge of cross-domain knowledge transferring and an Alignment Module to have stabilized recommendation results. Moreover, the task-oriented loss function further improves its performance. Based on extensive experiments with three real-world datasets, the results reveal the superiority of DiffCDR in a variety of cold-start and warm-start CDR tasks. Moreover, ablation experiments show that all the components of DiffCDR contribute critically to its performance.

%future work
This work explorers a new direction to tackle CDR challenges by DPMs with guidance. For the future, we will continue to emphasize the importance of the diffusion model architecture specially for CDR tasks. In addition, it is also meaningful to explorer the effects of other noise schedule for recommendation applications.

%------------------------------

\newpage
\bibliographystyle{ACM-Reference-Format}
\bibliography{sample-base}


\begin{thebibliography}{51}

%%% the Web address.

\ifx \showCODEN    \undefined \def \showCODEN     #1{\unskip}     \fi
\ifx \showDOI      \undefined \def \showDOI       #1{#1}\fi
\ifx \showISBNx    \undefined \def \showISBNx     #1{\unskip}     \fi
\ifx \showISBNxiii \undefined \def \showISBNxiii  #1{\unskip}     \fi
\ifx \showISSN     \undefined \def \showISSN      #1{\unskip}     \fi
\ifx \showLCCN     \undefined \def \showLCCN      #1{\unskip}     \fi
\ifx \shownote     \undefined \def \shownote      #1{#1}          \fi
\ifx \showarticletitle \undefined \def \showarticletitle #1{#1}   \fi
\ifx \showURL      \undefined \def \showURL       {\relax}        \fi
% invisible to TeX
\providecommand\bibfield[2]{#2}
\providecommand\bibinfo[2]{#2}
\providecommand\natexlab[1]{#1}
\providecommand\showeprint[2][]{arXiv:#2}

\bibitem[Bao et~al\mbox{.}(2022)]%
        {bao2022analytic}
\bibfield{author}{\bibinfo{person}{Fan Bao}, \bibinfo{person}{Chongxuan Li}, \bibinfo{person}{Jun Zhu}, {and} \bibinfo{person}{Bo Zhang}.} \bibinfo{year}{2022}\natexlab{}.
\newblock \showarticletitle{Analytic-dpm: an analytic estimate of the optimal reverse variance in diffusion probabilistic models}.
\newblock \bibinfo{journal}{\emph{arXiv preprint arXiv:2201.06503}} (\bibinfo{year}{2022}).
\newblock


\bibitem[Cao et~al\mbox{.}(2022)]%
        {cao2022cross}
\bibfield{author}{\bibinfo{person}{J. Cao}, \bibinfo{person}{J. Sheng}, \bibinfo{person}{X. Cong}, \bibinfo{person}{T. Liu}, {and} \bibinfo{person}{B. Wang}.} \bibinfo{year}{2022}\natexlab{}.
\newblock \showarticletitle{Cross-Domain Recommendation to Cold-Start Users via Variational Information Bottleneck}.
\newblock  (\bibinfo{date}{may} \bibinfo{year}{2022}), \bibinfo{pages}{2209--2223}.
\newblock
\urldef\tempurl%
\url{https://doi.org/10.1109/ICDE53745.2022.00211}
\showDOI{\tempurl}


\bibitem[Chen and de~Rijke(2018)]%
        {chen2018collective}
\bibfield{author}{\bibinfo{person}{Yifan Chen} {and} \bibinfo{person}{Maarten de Rijke}.} \bibinfo{year}{2018}\natexlab{}.
\newblock \showarticletitle{A collective variational autoencoder for top-n recommendation with side information}. In \bibinfo{booktitle}{\emph{Proceedings of the 3rd workshop on deep learning for recommender systems}}. \bibinfo{pages}{3--9}.
\newblock


\bibitem[Dhariwal and Nichol(2021)]%
        {dhariwal2021diffusion}
\bibfield{author}{\bibinfo{person}{Prafulla Dhariwal} {and} \bibinfo{person}{Alexander Nichol}.} \bibinfo{year}{2021}\natexlab{}.
\newblock \showarticletitle{Diffusion models beat gans on image synthesis}.
\newblock \bibinfo{journal}{\emph{Advances in Neural Information Processing Systems}}  \bibinfo{volume}{34} (\bibinfo{year}{2021}), \bibinfo{pages}{8780--8794}.
\newblock


\bibitem[Fu et~al\mbox{.}(2019)]%
        {10.1609/aaai.v33i01.330194}
\bibfield{author}{\bibinfo{person}{Wenjing Fu}, \bibinfo{person}{Zhaohui Peng}, \bibinfo{person}{Senzhang Wang}, \bibinfo{person}{Yang Xu}, {and} \bibinfo{person}{Jin Li}.} \bibinfo{year}{2019}\natexlab{}.
\newblock \showarticletitle{Deeply Fusing Reviews and Contents for Cold Start Users in Cross-Domain Recommendation Systems} \emph{(\bibinfo{series}{AAAI'19/IAAI'19/EAAI'19})}. \bibinfo{publisher}{AAAI Press}, Article \bibinfo{articleno}{12}, \bibinfo{numpages}{8}~pages.
\newblock
\showISBNx{978-1-57735-809-1}
\urldef\tempurl%
\url{https://doi.org/10.1609/aaai.v33i01.330194}
\showDOI{\tempurl}


\bibitem[Han et~al\mbox{.}(2022)]%
        {han2022card}
\bibfield{author}{\bibinfo{person}{Xizewen Han}, \bibinfo{person}{Huangjie Zheng}, {and} \bibinfo{person}{Mingyuan Zhou}.} \bibinfo{year}{2022}\natexlab{}.
\newblock \showarticletitle{CARD: Classification and regression diffusion models}.
\newblock \bibinfo{journal}{\emph{arXiv preprint arXiv:2206.07275}} (\bibinfo{year}{2022}).
\newblock


\bibitem[Ho et~al\mbox{.}(2020)]%
        {ho2020denoising}
\bibfield{author}{\bibinfo{person}{Jonathan Ho}, \bibinfo{person}{Ajay Jain}, {and} \bibinfo{person}{Pieter Abbeel}.} \bibinfo{year}{2020}\natexlab{}.
\newblock \showarticletitle{Denoising diffusion probabilistic models}.
\newblock \bibinfo{journal}{\emph{Advances in Neural Information Processing Systems}}  \bibinfo{volume}{33} (\bibinfo{year}{2020}), \bibinfo{pages}{6840--6851}.
\newblock


\bibitem[Ho and Salimans(2022)]%
        {ho2022classifier}
\bibfield{author}{\bibinfo{person}{Jonathan Ho} {and} \bibinfo{person}{Tim Salimans}.} \bibinfo{year}{2022}\natexlab{}.
\newblock \showarticletitle{Classifier-free diffusion guidance}.
\newblock \bibinfo{journal}{\emph{arXiv preprint arXiv:2207.12598}} (\bibinfo{year}{2022}).
\newblock


\bibitem[Kang et~al\mbox{.}(2019)]%
        {kang2019semi}
\bibfield{author}{\bibinfo{person}{SeongKu Kang}, \bibinfo{person}{Junyoung Hwang}, \bibinfo{person}{Dongha Lee}, {and} \bibinfo{person}{Hwanjo Yu}.} \bibinfo{year}{2019}\natexlab{}.
\newblock \showarticletitle{Semi-supervised learning for cross-domain recommendation to cold-start users}. In \bibinfo{booktitle}{\emph{Proceedings of the 28th ACM International Conference on Information and Knowledge Management}}. \bibinfo{pages}{1563--1572}.
\newblock


\bibitem[Kingma et~al\mbox{.}(2021)]%
        {kingma2021variational}
\bibfield{author}{\bibinfo{person}{Diederik Kingma}, \bibinfo{person}{Tim Salimans}, \bibinfo{person}{Ben Poole}, {and} \bibinfo{person}{Jonathan Ho}.} \bibinfo{year}{2021}\natexlab{}.
\newblock \showarticletitle{Variational diffusion models}.
\newblock \bibinfo{journal}{\emph{Advances in neural information processing systems}}  \bibinfo{volume}{34} (\bibinfo{year}{2021}), \bibinfo{pages}{21696--21707}.
\newblock


\bibitem[Kong et~al\mbox{.}(2020)]%
        {kong2020diffwave}
\bibfield{author}{\bibinfo{person}{Zhifeng Kong}, \bibinfo{person}{Wei Ping}, \bibinfo{person}{Jiaji Huang}, \bibinfo{person}{Kexin Zhao}, {and} \bibinfo{person}{Bryan Catanzaro}.} \bibinfo{year}{2020}\natexlab{}.
\newblock \showarticletitle{Diffwave: A versatile diffusion model for audio synthesis}.
\newblock \bibinfo{journal}{\emph{arXiv preprint arXiv:2009.09761}} (\bibinfo{year}{2020}).
\newblock


\bibitem[Koren et~al\mbox{.}(2009)]%
        {koren2009matrix}
\bibfield{author}{\bibinfo{person}{Yehuda Koren}, \bibinfo{person}{Robert Bell}, {and} \bibinfo{person}{Chris Volinsky}.} \bibinfo{year}{2009}\natexlab{}.
\newblock \showarticletitle{Matrix factorization techniques for recommender systems}.
\newblock \bibinfo{journal}{\emph{Computer}} \bibinfo{volume}{42}, \bibinfo{number}{8} (\bibinfo{year}{2009}), \bibinfo{pages}{30--37}.
\newblock


\bibitem[Krishnan et~al\mbox{.}(2020)]%
        {krishnan2020transfer}
\bibfield{author}{\bibinfo{person}{Adit Krishnan}, \bibinfo{person}{Mahashweta Das}, \bibinfo{person}{Mangesh Bendre}, \bibinfo{person}{Hao Yang}, {and} \bibinfo{person}{Hari Sundaram}.} \bibinfo{year}{2020}\natexlab{}.
\newblock \showarticletitle{Transfer learning via contextual invariants for one-to-many cross-domain recommendation}. In \bibinfo{booktitle}{\emph{Proceedings of the 43rd International ACM SIGIR Conference on Research and Development in Information Retrieval}}. \bibinfo{pages}{1081--1090}.
\newblock


\bibitem[Lee et~al\mbox{.}(2019)]%
        {lee2019melu}
\bibfield{author}{\bibinfo{person}{Hoyeop Lee}, \bibinfo{person}{Jinbae Im}, \bibinfo{person}{Seongwon Jang}, \bibinfo{person}{Hyunsouk Cho}, {and} \bibinfo{person}{Sehee Chung}.} \bibinfo{year}{2019}\natexlab{}.
\newblock \showarticletitle{Melu: Meta-learned user preference estimator for cold-start recommendation}. In \bibinfo{booktitle}{\emph{Proceedings of the 25th ACM SIGKDD International Conference on Knowledge Discovery \& Data Mining}}. \bibinfo{pages}{1073--1082}.
\newblock


\bibitem[Li et~al\mbox{.}(2022)]%
        {li2022diffusion}
\bibfield{author}{\bibinfo{person}{Xiang~Lisa Li}, \bibinfo{person}{John Thickstun}, \bibinfo{person}{Ishaan Gulrajani}, \bibinfo{person}{Percy Liang}, {and} \bibinfo{person}{Tatsunori~B Hashimoto}.} \bibinfo{year}{2022}\natexlab{}.
\newblock \showarticletitle{Diffusion-LM Improves Controllable Text Generation}.
\newblock \bibinfo{journal}{\emph{arXiv preprint arXiv:2205.14217}} (\bibinfo{year}{2022}).
\newblock


\bibitem[Liang et~al\mbox{.}(2018)]%
        {liang2018variational}
\bibfield{author}{\bibinfo{person}{Dawen Liang}, \bibinfo{person}{Rahul~G Krishnan}, \bibinfo{person}{Matthew~D Hoffman}, {and} \bibinfo{person}{Tony Jebara}.} \bibinfo{year}{2018}\natexlab{}.
\newblock \showarticletitle{Variational autoencoders for collaborative filtering}. In \bibinfo{booktitle}{\emph{Proceedings of the 2018 world wide web conference}}. \bibinfo{pages}{689--698}.
\newblock


\bibitem[Liu et~al\mbox{.}(2022)]%
        {liu2022exploiting}
\bibfield{author}{\bibinfo{person}{Weiming Liu}, \bibinfo{person}{Xiaolin Zheng}, \bibinfo{person}{Jiajie Su}, \bibinfo{person}{Mengling Hu}, \bibinfo{person}{Yanchao Tan}, {and} \bibinfo{person}{Chaochao Chen}.} \bibinfo{year}{2022}\natexlab{}.
\newblock \showarticletitle{Exploiting variational domain-invariant user embedding for partially overlapped cross domain recommendation}. In \bibinfo{booktitle}{\emph{Proceedings of the 45th International ACM SIGIR Conference on Research and Development in Information Retrieval}}. \bibinfo{pages}{312--321}.
\newblock


\bibitem[Lu et~al\mbox{.}(2022a)]%
        {lu2022dpm}
\bibfield{author}{\bibinfo{person}{Cheng Lu}, \bibinfo{person}{Yuhao Zhou}, \bibinfo{person}{Fan Bao}, \bibinfo{person}{Jianfei Chen}, \bibinfo{person}{Chongxuan Li}, {and} \bibinfo{person}{Jun Zhu}.} \bibinfo{year}{2022}\natexlab{a}.
\newblock \showarticletitle{DPM-Solver: A Fast ODE Solver for Diffusion Probabilistic Model Sampling in Around 10 Steps}.
\newblock \bibinfo{journal}{\emph{arXiv preprint arXiv:2206.00927}} (\bibinfo{year}{2022}).
\newblock


\bibitem[Lu et~al\mbox{.}(2022b)]%
        {lu2022dpmsolver}
\bibfield{author}{\bibinfo{person}{Cheng Lu}, \bibinfo{person}{Yuhao Zhou}, \bibinfo{person}{Fan Bao}, \bibinfo{person}{Jianfei Chen}, \bibinfo{person}{Chongxuan Li}, {and} \bibinfo{person}{Jun Zhu}.} \bibinfo{year}{2022}\natexlab{b}.
\newblock \bibinfo{title}{DPM-Solver++: Fast Solver for Guided Sampling of Diffusion Probabilistic Models}.
\newblock
\newblock
\showeprint[arxiv]{2211.01095}~[cs.LG]


\bibitem[Ma et~al\mbox{.}(2019)]%
        {10.1145/3331184.3331200}
\bibfield{author}{\bibinfo{person}{Muyang Ma}, \bibinfo{person}{Pengjie Ren}, \bibinfo{person}{Yujie Lin}, \bibinfo{person}{Zhumin Chen}, \bibinfo{person}{Jun Ma}, {and} \bibinfo{person}{Maarten~de Rijke}.} \bibinfo{year}{2019}\natexlab{}.
\newblock \showarticletitle{$\pi$-Net: A Parallel Information-Sharing Network for Shared-Account Cross-Domain Sequential Recommendations}. In \bibinfo{booktitle}{\emph{Proceedings of the 42nd International ACM SIGIR Conference on Research and Development in Information Retrieval}} (Paris, France) \emph{(\bibinfo{series}{SIGIR'19})}. \bibinfo{publisher}{Association for Computing Machinery}, \bibinfo{address}{New York, NY, USA}, \bibinfo{pages}{685–694}.
\newblock
\showISBNx{9781450361729}
\urldef\tempurl%
\url{https://doi.org/10.1145/3331184.3331200}
\showDOI{\tempurl}


\bibitem[Man et~al\mbox{.}(2017)]%
        {man2017cross}
\bibfield{author}{\bibinfo{person}{Tong Man}, \bibinfo{person}{Huawei Shen}, \bibinfo{person}{Xiaolong Jin}, {and} \bibinfo{person}{Xueqi Cheng}.} \bibinfo{year}{2017}\natexlab{}.
\newblock \showarticletitle{Cross-domain recommendation: An embedding and mapping approach.}. In \bibinfo{booktitle}{\emph{IJCAI}}, Vol.~\bibinfo{volume}{17}. \bibinfo{pages}{2464--2470}.
\newblock


\bibitem[Munkhdalai and Yu(2017)]%
        {munkhdalai2017meta}
\bibfield{author}{\bibinfo{person}{Tsendsuren Munkhdalai} {and} \bibinfo{person}{Hong Yu}.} \bibinfo{year}{2017}\natexlab{}.
\newblock \showarticletitle{Meta networks}. In \bibinfo{booktitle}{\emph{International Conference on Machine Learning}}. PMLR, \bibinfo{pages}{2554--2563}.
\newblock


\bibitem[Nichol et~al\mbox{.}(2021)]%
        {nichol2021glide}
\bibfield{author}{\bibinfo{person}{Alex Nichol}, \bibinfo{person}{Prafulla Dhariwal}, \bibinfo{person}{Aditya Ramesh}, \bibinfo{person}{Pranav Shyam}, \bibinfo{person}{Pamela Mishkin}, \bibinfo{person}{Bob McGrew}, \bibinfo{person}{Ilya Sutskever}, {and} \bibinfo{person}{Mark Chen}.} \bibinfo{year}{2021}\natexlab{}.
\newblock \showarticletitle{Glide: Towards photorealistic image generation and editing with text-guided diffusion models}.
\newblock \bibinfo{journal}{\emph{arXiv preprint arXiv:2112.10741}} (\bibinfo{year}{2021}).
\newblock


\bibitem[Pan et~al\mbox{.}(2019)]%
        {pan2019warm}
\bibfield{author}{\bibinfo{person}{Feiyang Pan}, \bibinfo{person}{Shuokai Li}, \bibinfo{person}{Xiang Ao}, \bibinfo{person}{Pingzhong Tang}, {and} \bibinfo{person}{Qing He}.} \bibinfo{year}{2019}\natexlab{}.
\newblock \showarticletitle{Warm up cold-start advertisements: Improving ctr predictions via learning to learn id embeddings}. In \bibinfo{booktitle}{\emph{Proceedings of the 42nd International ACM SIGIR Conference on Research and Development in Information Retrieval}}. \bibinfo{pages}{695--704}.
\newblock


\bibitem[Popov et~al\mbox{.}(2021)]%
        {popov2021diffusion}
\bibfield{author}{\bibinfo{person}{Vadim Popov}, \bibinfo{person}{Ivan Vovk}, \bibinfo{person}{Vladimir Gogoryan}, \bibinfo{person}{Tasnima Sadekova}, \bibinfo{person}{Mikhail Kudinov}, {and} \bibinfo{person}{Jiansheng Wei}.} \bibinfo{year}{2021}\natexlab{}.
\newblock \showarticletitle{Diffusion-based voice conversion with fast maximum likelihood sampling scheme}.
\newblock \bibinfo{journal}{\emph{arXiv preprint arXiv:2109.13821}} (\bibinfo{year}{2021}).
\newblock


\bibitem[Ramesh et~al\mbox{.}(2022)]%
        {ramesh2022hierarchical}
\bibfield{author}{\bibinfo{person}{Aditya Ramesh}, \bibinfo{person}{Prafulla Dhariwal}, \bibinfo{person}{Alex Nichol}, \bibinfo{person}{Casey Chu}, {and} \bibinfo{person}{Mark Chen}.} \bibinfo{year}{2022}\natexlab{}.
\newblock \showarticletitle{Hierarchical text-conditional image generation with clip latents}.
\newblock \bibinfo{journal}{\emph{arXiv preprint arXiv:2204.06125}} (\bibinfo{year}{2022}).
\newblock


\bibitem[Ramesh et~al\mbox{.}(2021)]%
        {ramesh2021zero}
\bibfield{author}{\bibinfo{person}{Aditya Ramesh}, \bibinfo{person}{Mikhail Pavlov}, \bibinfo{person}{Gabriel Goh}, \bibinfo{person}{Scott Gray}, \bibinfo{person}{Chelsea Voss}, \bibinfo{person}{Alec Radford}, \bibinfo{person}{Mark Chen}, {and} \bibinfo{person}{Ilya Sutskever}.} \bibinfo{year}{2021}\natexlab{}.
\newblock \showarticletitle{Zero-shot text-to-image generation}. In \bibinfo{booktitle}{\emph{International Conference on Machine Learning}}. PMLR, \bibinfo{pages}{8821--8831}.
\newblock


\bibitem[Rendle et~al\mbox{.}(2012)]%
        {rendle2012bpr}
\bibfield{author}{\bibinfo{person}{Steffen Rendle}, \bibinfo{person}{Christoph Freudenthaler}, \bibinfo{person}{Zeno Gantner}, {and} \bibinfo{person}{Lars Schmidt-Thieme}.} \bibinfo{year}{2012}\natexlab{}.
\newblock \showarticletitle{BPR: Bayesian personalized ranking from implicit feedback}.
\newblock \bibinfo{journal}{\emph{arXiv preprint arXiv:1205.2618}} (\bibinfo{year}{2012}).
\newblock


\bibitem[Ricci et~al\mbox{.}(2011)]%
        {ricci2011introduction}
\bibfield{author}{\bibinfo{person}{Francesco Ricci}, \bibinfo{person}{Lior Rokach}, {and} \bibinfo{person}{Bracha Shapira}.} \bibinfo{year}{2011}\natexlab{}.
\newblock \showarticletitle{Introduction to recommender systems handbook}.
\newblock In \bibinfo{booktitle}{\emph{Recommender systems handbook}}. \bibinfo{publisher}{Springer}, \bibinfo{pages}{1--35}.
\newblock


\bibitem[Rombach et~al\mbox{.}(2022)]%
        {rombach2022high}
\bibfield{author}{\bibinfo{person}{Robin Rombach}, \bibinfo{person}{Andreas Blattmann}, \bibinfo{person}{Dominik Lorenz}, \bibinfo{person}{Patrick Esser}, {and} \bibinfo{person}{Bj{\"o}rn Ommer}.} \bibinfo{year}{2022}\natexlab{}.
\newblock \showarticletitle{High-resolution image synthesis with latent diffusion models}. In \bibinfo{booktitle}{\emph{Proceedings of the IEEE/CVF Conference on Computer Vision and Pattern Recognition}}. \bibinfo{pages}{10684--10695}.
\newblock


\bibitem[Ronneberger et~al\mbox{.}(2015)]%
        {ronneberger2015u}
\bibfield{author}{\bibinfo{person}{Olaf Ronneberger}, \bibinfo{person}{Philipp Fischer}, {and} \bibinfo{person}{Thomas Brox}.} \bibinfo{year}{2015}\natexlab{}.
\newblock \showarticletitle{U-net: Convolutional networks for biomedical image segmentation}. In \bibinfo{booktitle}{\emph{International Conference on Medical image computing and computer-assisted intervention}}. Springer, \bibinfo{pages}{234--241}.
\newblock


\bibitem[Salah et~al\mbox{.}(2021)]%
        {salah2021towards}
\bibfield{author}{\bibinfo{person}{Aghiles Salah}, \bibinfo{person}{Thanh~Binh Tran}, {and} \bibinfo{person}{Hady Lauw}.} \bibinfo{year}{2021}\natexlab{}.
\newblock \showarticletitle{Towards Source-Aligned Variational Models for Cross-Domain Recommendation}. In \bibinfo{booktitle}{\emph{Fifteenth ACM Conference on Recommender Systems}}. \bibinfo{pages}{176--186}.
\newblock


\bibitem[Salimans and Ho(2022)]%
        {salimans2022progressive}
\bibfield{author}{\bibinfo{person}{Tim Salimans} {and} \bibinfo{person}{Jonathan Ho}.} \bibinfo{year}{2022}\natexlab{}.
\newblock \showarticletitle{Progressive distillation for fast sampling of diffusion models}.
\newblock \bibinfo{journal}{\emph{arXiv preprint arXiv:2202.00512}} (\bibinfo{year}{2022}).
\newblock


\bibitem[Shenbin et~al\mbox{.}(2020)]%
        {shenbin2020recvae}
\bibfield{author}{\bibinfo{person}{Ilya Shenbin}, \bibinfo{person}{Anton Alekseev}, \bibinfo{person}{Elena Tutubalina}, \bibinfo{person}{Valentin Malykh}, {and} \bibinfo{person}{Sergey~I Nikolenko}.} \bibinfo{year}{2020}\natexlab{}.
\newblock \showarticletitle{Recvae: A new variational autoencoder for top-n recommendations with implicit feedback}. In \bibinfo{booktitle}{\emph{Proceedings of the 13th International Conference on Web Search and Data Mining}}. \bibinfo{pages}{528--536}.
\newblock


\bibitem[Singh and Gordon(2008)]%
        {singh2008relational}
\bibfield{author}{\bibinfo{person}{Ajit~P Singh} {and} \bibinfo{person}{Geoffrey~J Gordon}.} \bibinfo{year}{2008}\natexlab{}.
\newblock \showarticletitle{Relational learning via collective matrix factorization}. In \bibinfo{booktitle}{\emph{Proceedings of the 14th ACM SIGKDD international conference on Knowledge discovery and data mining}}. \bibinfo{pages}{650--658}.
\newblock


\bibitem[Sohl-Dickstein et~al\mbox{.}(2015)]%
        {sohl2015deep}
\bibfield{author}{\bibinfo{person}{Jascha Sohl-Dickstein}, \bibinfo{person}{Eric Weiss}, \bibinfo{person}{Niru Maheswaranathan}, {and} \bibinfo{person}{Surya Ganguli}.} \bibinfo{year}{2015}\natexlab{}.
\newblock \showarticletitle{Deep unsupervised learning using nonequilibrium thermodynamics}. In \bibinfo{booktitle}{\emph{International Conference on Machine Learning}}. PMLR, \bibinfo{pages}{2256--2265}.
\newblock


\bibitem[Song et~al\mbox{.}(2020a)]%
        {song2020denoising}
\bibfield{author}{\bibinfo{person}{Jiaming Song}, \bibinfo{person}{Chenlin Meng}, {and} \bibinfo{person}{Stefano Ermon}.} \bibinfo{year}{2020}\natexlab{a}.
\newblock \showarticletitle{Denoising diffusion implicit models}.
\newblock \bibinfo{journal}{\emph{arXiv preprint arXiv:2010.02502}} (\bibinfo{year}{2020}).
\newblock


\bibitem[Song et~al\mbox{.}(2020b)]%
        {song2020score}
\bibfield{author}{\bibinfo{person}{Yang Song}, \bibinfo{person}{Jascha Sohl-Dickstein}, \bibinfo{person}{Diederik~P Kingma}, \bibinfo{person}{Abhishek Kumar}, \bibinfo{person}{Stefano Ermon}, {and} \bibinfo{person}{Ben Poole}.} \bibinfo{year}{2020}\natexlab{b}.
\newblock \showarticletitle{Score-based generative modeling through stochastic differential equations}.
\newblock \bibinfo{journal}{\emph{arXiv preprint arXiv:2011.13456}} (\bibinfo{year}{2020}).
\newblock


\bibitem[Truong et~al\mbox{.}(2021)]%
        {truong2021bilateral}
\bibfield{author}{\bibinfo{person}{Quoc-Tuan Truong}, \bibinfo{person}{Aghiles Salah}, {and} \bibinfo{person}{Hady~W Lauw}.} \bibinfo{year}{2021}\natexlab{}.
\newblock \showarticletitle{Bilateral variational autoencoder for collaborative filtering}. In \bibinfo{booktitle}{\emph{Proceedings of the 14th ACM International Conference on Web Search and Data Mining}}. \bibinfo{pages}{292--300}.
\newblock


\bibitem[Van~der Maaten and Hinton(2008)]%
        {van2008visualizing}
\bibfield{author}{\bibinfo{person}{Laurens Van~der Maaten} {and} \bibinfo{person}{Geoffrey Hinton}.} \bibinfo{year}{2008}\natexlab{}.
\newblock \showarticletitle{Visualizing data using t-SNE.}
\newblock \bibinfo{journal}{\emph{Journal of machine learning research}} \bibinfo{volume}{9}, \bibinfo{number}{11} (\bibinfo{year}{2008}).
\newblock


\bibitem[Vaswani et~al\mbox{.}(2017)]%
        {vaswani2017attention}
\bibfield{author}{\bibinfo{person}{Ashish Vaswani}, \bibinfo{person}{Noam Shazeer}, \bibinfo{person}{Niki Parmar}, \bibinfo{person}{Jakob Uszkoreit}, \bibinfo{person}{Llion Jones}, \bibinfo{person}{Aidan~N Gomez}, \bibinfo{person}{{\L}ukasz Kaiser}, {and} \bibinfo{person}{Illia Polosukhin}.} \bibinfo{year}{2017}\natexlab{}.
\newblock \showarticletitle{Attention is all you need}.
\newblock \bibinfo{journal}{\emph{Advances in neural information processing systems}}  \bibinfo{volume}{30} (\bibinfo{year}{2017}).
\newblock


\bibitem[Wang et~al\mbox{.}(2021)]%
        {10.1145/3459637.3482137}
\bibfield{author}{\bibinfo{person}{Tianxin Wang}, \bibinfo{person}{Fuzhen Zhuang}, \bibinfo{person}{Zhiqiang Zhang}, \bibinfo{person}{Daixin Wang}, \bibinfo{person}{Jun Zhou}, {and} \bibinfo{person}{Qing He}.} \bibinfo{year}{2021}\natexlab{}.
\newblock \showarticletitle{Low-Dimensional Alignment for Cross-Domain Recommendation} \emph{(\bibinfo{series}{CIKM '21})}. \bibinfo{publisher}{Association for Computing Machinery}, \bibinfo{address}{New York, NY, USA}, \bibinfo{pages}{3508–3512}.
\newblock
\showISBNx{9781450384469}
\urldef\tempurl%
\url{https://doi.org/10.1145/3459637.3482137}
\showURL{%
\tempurl}


\bibitem[Wang et~al\mbox{.}(2023)]%
        {wang2023diffusion}
\bibfield{author}{\bibinfo{person}{Wenjie Wang}, \bibinfo{person}{Yiyan Xu}, \bibinfo{person}{Fuli Feng}, \bibinfo{person}{Xinyu Lin}, \bibinfo{person}{Xiangnan He}, {and} \bibinfo{person}{Tat-Seng Chua}.} \bibinfo{year}{2023}\natexlab{}.
\newblock \bibinfo{title}{Diffusion Recommender Model}.
\newblock
\newblock
\showeprint[arxiv]{2304.04971}~[cs.IR]


\bibitem[Wu et~al\mbox{.}(2022)]%
        {wu2022dynamics}
\bibfield{author}{\bibinfo{person}{Junda Wu}, \bibinfo{person}{Zhihui Xie}, \bibinfo{person}{Tong Yu}, \bibinfo{person}{Handong Zhao}, \bibinfo{person}{Ruiyi Zhang}, {and} \bibinfo{person}{Shuai Li}.} \bibinfo{year}{2022}\natexlab{}.
\newblock \showarticletitle{Dynamics-Aware Adaptation for Reinforcement Learning Based Cross-Domain Interactive Recommendation}. In \bibinfo{booktitle}{\emph{Proceedings of the 45th International ACM SIGIR Conference on Research and Development in Information Retrieval}}. \bibinfo{pages}{290--300}.
\newblock


\bibitem[Zang et~al\mbox{.}(2021)]%
        {DBLP:journals/corr/abs-2108-03357}
\bibfield{author}{\bibinfo{person}{Tianzi Zang}, \bibinfo{person}{Yanmin Zhu}, \bibinfo{person}{Haobing Liu}, \bibinfo{person}{Ruohan Zhang}, {and} \bibinfo{person}{Jiadi Yu}.} \bibinfo{year}{2021}\natexlab{}.
\newblock \showarticletitle{A Survey on Cross-domain Recommendation: Taxonomies, Methods, and Future Directions}.
\newblock \bibinfo{journal}{\emph{CoRR}}  \bibinfo{volume}{abs/2108.03357} (\bibinfo{year}{2021}).
\newblock
\showeprint[arXiv]{2108.03357}
\urldef\tempurl%
\url{https://arxiv.org/abs/2108.03357}
\showURL{%
\tempurl}


\bibitem[Zhao et~al\mbox{.}(2020)]%
        {zhao2020catn}
\bibfield{author}{\bibinfo{person}{Cheng Zhao}, \bibinfo{person}{Chenliang Li}, \bibinfo{person}{Rong Xiao}, \bibinfo{person}{Hongbo Deng}, {and} \bibinfo{person}{Aixin Sun}.} \bibinfo{year}{2020}\natexlab{}.
\newblock \showarticletitle{CATN: Cross-domain recommendation for cold-start users via aspect transfer network}. In \bibinfo{booktitle}{\emph{Proceedings of the 43rd International ACM SIGIR Conference on Research and Development in Information Retrieval}}. \bibinfo{pages}{229--238}.
\newblock


\bibitem[Zhu et~al\mbox{.}(2020)]%
        {zhu2020deep}
\bibfield{author}{\bibinfo{person}{Feng Zhu}, \bibinfo{person}{Yan Wang}, \bibinfo{person}{Chaochao Chen}, \bibinfo{person}{Guanfeng Liu}, \bibinfo{person}{Mehmet Orgun}, {and} \bibinfo{person}{Jia Wu}.} \bibinfo{year}{2020}\natexlab{}.
\newblock \showarticletitle{A deep framework for cross-domain and cross-system recommendations}.
\newblock \bibinfo{journal}{\emph{arXiv preprint arXiv:2009.06215}} (\bibinfo{year}{2020}).
\newblock


\bibitem[Zhu et~al\mbox{.}(2021b)]%
        {zhu2021cross}
\bibfield{author}{\bibinfo{person}{Feng Zhu}, \bibinfo{person}{Yan Wang}, \bibinfo{person}{Chaochao Chen}, \bibinfo{person}{Jun Zhou}, \bibinfo{person}{Longfei Li}, {and} \bibinfo{person}{Guanfeng Liu}.} \bibinfo{year}{2021}\natexlab{b}.
\newblock \showarticletitle{Cross-domain recommendation: challenges, progress, and prospects}.
\newblock \bibinfo{journal}{\emph{arXiv preprint arXiv:2103.01696}} (\bibinfo{year}{2021}).
\newblock


\bibitem[Zhu et~al\mbox{.}(2021a)]%
        {zhu2021transfer}
\bibfield{author}{\bibinfo{person}{Yongchun Zhu}, \bibinfo{person}{Kaikai Ge}, \bibinfo{person}{Fuzhen Zhuang}, \bibinfo{person}{Ruobing Xie}, \bibinfo{person}{Dongbo Xi}, \bibinfo{person}{Xu Zhang}, \bibinfo{person}{Leyu Lin}, {and} \bibinfo{person}{Qing He}.} \bibinfo{year}{2021}\natexlab{a}.
\newblock \showarticletitle{Transfer-meta framework for cross-domain recommendation to cold-start users}. In \bibinfo{booktitle}{\emph{Proceedings of the 44th International ACM SIGIR Conference on Research and Development in Information Retrieval}}. \bibinfo{pages}{1813--1817}.
\newblock


\bibitem[Zhu et~al\mbox{.}(2022)]%
        {zhu2022personalized}
\bibfield{author}{\bibinfo{person}{Yongchun Zhu}, \bibinfo{person}{Zhenwei Tang}, \bibinfo{person}{Yudan Liu}, \bibinfo{person}{Fuzhen Zhuang}, \bibinfo{person}{Ruobing Xie}, \bibinfo{person}{Xu Zhang}, \bibinfo{person}{Leyu Lin}, {and} \bibinfo{person}{Qing He}.} \bibinfo{year}{2022}\natexlab{}.
\newblock \showarticletitle{Personalized transfer of user preferences for cross-domain recommendation}. In \bibinfo{booktitle}{\emph{Proceedings of the Fifteenth ACM International Conference on Web Search and Data Mining}}. \bibinfo{pages}{1507--1515}.
\newblock


\bibitem[Zhu et~al\mbox{.}(2021c)]%
        {zhu2021learning}
\bibfield{author}{\bibinfo{person}{Yongchun Zhu}, \bibinfo{person}{Ruobing Xie}, \bibinfo{person}{Fuzhen Zhuang}, \bibinfo{person}{Kaikai Ge}, \bibinfo{person}{Ying Sun}, \bibinfo{person}{Xu Zhang}, \bibinfo{person}{Leyu Lin}, {and} \bibinfo{person}{Juan Cao}.} \bibinfo{year}{2021}\natexlab{c}.
\newblock \showarticletitle{Learning to warm up cold item embeddings for cold-start recommendation with meta scaling and shifting networks}. In \bibinfo{booktitle}{\emph{Proceedings of the 44th International ACM SIGIR Conference on Research and Development in Information Retrieval}}. \bibinfo{pages}{1167--1176}.
\newblock


\end{thebibliography}

\appendix

\section{Background of DPMs}
%forward,backward
%DPMs are powerful generative models, they have shown impressive performance in image synthesis  \cite{ramesh2022hierarchical,rombach2022high} and other generative tasks  \cite{kong2020diffwave,li2022diffusion} recently.
 \label{sec:add}
DPMs define a forward process $\{ x_t | t \in [0,T] , T>0\} $ given $x_0 \sim q(x_{0})$. Noise will be add to the data, and the distribution of $x_t$ conditioned on  $x_{t-1}$ is defined as$ q( x_t | x_{t-1}) = \mathcal{N}(x_t| \sqrt{ \alpha_t}x_{t-1}, (1-\alpha_t)\mathcal{I}).$ 
If the noise $1-\alpha_t$ in each time step is small enough, the reverse process $ q'( x_{t-1} | x_{t} ) $ can be approximated by Gaussian distribution too \cite{ ho2020denoising,nichol2021glide}. Then we can have a model to learn this distribution by $ p_\theta ( x_{t-1}|x_t) = \mathcal{N}( \mu_{\theta}, \Sigma_{\theta} ).$
If the total noise $\{ (\sqrt{\alpha_1},(1-\alpha_1)), (\sqrt{\alpha_2},(1-\alpha_2)), ...\}$ is large enough, then the state $ x_T $ will be close to $ \mathcal{N} \sim (0,\mathcal{I})$. Now we can start from $ x_T $ and use $ p_\theta $ iteratively to produce  $ x_T-1,...,x_0 $.

%generate samples q(xt|x0) by gaussian (cite),used to train p_\theta
It is possible to generate samples $x_t \sim q(x_t | x_0)$ with arbitrary $t$ by adding Gaussian noise to $x_0$, that gives
$$q(x_t | x_0) = \mathcal{N}( x_t ; \bar{\alpha}(t)x_0, \sigma^{2}(t)\mathcal{I}).  $$ Here $ \bar{\alpha}(t)$ and $\sigma(t)$ is noise schedule of DPM, and let them denote as $ \bar{\alpha}_t , \sigma_t$ respectively for neatness.

%loss definie (cite)
With these noised samples, we can train a function $\epsilon_\theta $ to approximate the added noise, 
we have its loss from the work \cite{ho2020denoising} as:
\begin{equation}
\mathcal{L}_{simple}(\theta) = \mathbb{E}_{t \in [0,T],x_{0}\sim q(x_0), \epsilon \sim\mathcal{N}(0,\mathcal{I}) } \left[  || \epsilon - \epsilon_{\theta}( x_t , t)  ||^2 \right]
 \label{diff_loss}
\end{equation}
where $\epsilon \sim \mathcal{N}(0,\mathcal{I})$.
The authors also show that $\mu_{\theta}$ can be derived from $\epsilon_\theta$ with constant $\Sigma_{\theta}$.

\subsection{Guided DPMs}
%guided diffusion with classifier 
With gradients from a classifier $p_{\phi}(y|x_t)$, Dhariwal et al.  \cite{dhariwal2021diffusion} propose a conditional DPM. They define a new noise approximate function $ \hat{\epsilon}_{\theta}(x_t) = \epsilon_{\theta} - s \cdot \Sigma_{\theta} \nabla_{x_t} \log p_{\phi}(y|x_t)$.
Here $s$ is the gradient scale. It is claimed that a larger value of $s$ can produce higher accurate but less diverse samples.

%classifier-free
A recent work \cite{ho2022classifier} proposes a classifier-free guided DPM, which does not require an additional classifier to provide gradient by a conditional distribution $p_{\phi}(y|x_t)$. In this work, authors replace $\log p_{\phi}(y|x_t)$ by  $\epsilon_{\theta}( x_t | y)$ and $\epsilon_{\theta}( x_t )$, and derive the noise approximate function without an extra classifier:
\begin{equation}
 \hat{\epsilon}_{\theta}(x_t|y) = \epsilon_{\theta}(x_t) + s \cdot (\epsilon_{\theta}(x_t|y) - \epsilon_{\theta}(x_t))    
\label{class_free}
\end{equation}
By Bayes rule, an implicit classifier $p^{i}_{\phi}(y|x_t) \propto \frac{p(x_t|y)}{ p(x_t)}$, and we can have $ \nabla_{x_t} \log  p^{i}_{\phi}(y|x_t) \propto  \nabla_{x_t} \log p(x_t|y) -  \nabla_{x_t} \log p(x_t) =$ \\
$  - \frac{1}{\Sigma_{\theta}} \left( \epsilon^{*}_{\theta}(x_t|y) - \epsilon^{*}_{\theta}(x_t) \right)$, $\epsilon^{*}_{\theta} $ is the truth score function.  $\frac{1}{\Sigma_{\theta}}$ can be viewed as a part of $s$ if we set it constant.

\subsection{Sampling Acceleration}
%shortage of diffusion model
DMPs generate samples according to the reverse process $ q'( x_{t-1} | x_{t} ) $, thus it usually requires hundreds of steps to produce predictions. It is time-consuming compared to other generative models with VAE and GAN frameworks, which may limit its application in high-throughput scenarios. Efficient samplers have become a central research topic in DMP.

%SDE 2.2
Before introducing the fast diffusion solvers we adopt, let us give a brief about diffusion ordinary differential equation (ODE).  
Kingma et al.  \cite{kingma2021variational} prove that the stochastic differential equation (SDE) $ {\rm d} x_{t} = f(t)x_{t}{\rm d} t + g(t){\rm d}w_{t} $ has the same transition probability as $q( x_t | x_{t-1})$, where $w_{t}$ is standard Wiener process and $f(t) =   \frac{{\rm d \, log} \bar{\alpha}_t} { {\rm d}t} , g^2(t) = \frac{{\rm d} \sigma^2_t}{{\rm d}t} - 2 \frac{{\rm d \, log} \bar{\alpha}_t}{{\rm d}t}\sigma^2_t$.

%reverse of SDE 2.4
Song et al. \cite{song2020score} shows that a reverse process $[T,0]$ is equivalent to the forward process:
$ {\rm d} x_t = [f(t)x_t - g^2(t) \nabla_x {\rm log} q_t(x_t)]{\rm d}t + g(t){\rm d} \bar {w}_t$, where $\bar {w}_t$ is standard Wiener process.
%reverse of ODE 2.6,2.7
They also prove the probability flow ODE is 
$ \frac{ {\rm d} x_t }{ {\rm d }t} = f(t)x_t  -  \frac{1}{2}  g^2(t) \nabla_x {\rm log}  q_t(x_t)  $, and they define the diffusion ODE as
\begin{equation}
\frac{ {\rm d} x_t }{ {\rm d }t} = f(t)x_t + \frac{g^2(t)}{2\sigma_t}   \epsilon_{\theta}( x_t, t).
\label{diff_ode}
\end{equation}
DPM-Solver  \cite{lu2022dpmsolver}  is a fast training-free sampling method for DPMs. It takes advantage of the semi-linear structure of diffusion ODE and can produce comparable results within only 20 function evaluations. The authors of this work observe that the  RHS of diffusion ODE Equation (\ref{diff_ode}) consists of  a linear and a nonlinear function of $x_t$, and this structure is referred to as semi-linear ODE. Whereas previous works treat the two terms together, this approach treats the two parts separately, which can reduce the approximation error. 
Given the reverse process with $M+1$ time steps 
$\{ t_i \}^M_{i=0}$ from $t_0 = T$ to $t_M = 0$, and the initial value $ \tilde{x}_{t_0} = x_T$,  they simplify the nonlinear part and define the DPM-solver-1 as $ \tilde{x}_{t_i} = \frac{\alpha_{t_i}}{\alpha_{t_{i-1}}} \tilde{x}_{t_{i-1}} - \sigma_{t_i} ( e^{h_i}-1) \epsilon_{\theta} (\tilde{x}_{t_{i-1}}),$
where $\lambda_{t_i} = {\rm log} \left( \frac{\alpha_{t_i} }{\sigma_{t_i}} \right) $,
$h_i = \lambda_{t_i} - \lambda_{t_{i-1}}$. In addition, they propose higher-order solvers that are claimed to converge with fewer inference steps.

\section{Ethical Considerations}
As a recommender model, our DiffCDR is able to improve user experience and satisfaction by leveraging behavioral records. 
However, recommender systems may also create filter bubbles that limit the diversity of content exposed to users. As a result, its application may introduce stereotypes or amplify inequalities, and lead to biased recommendation results that may bring negative influence on certain individuals or groups in society.
In order to train the recommender model, the user's profile and behavioral records are collected. During this process, the user's privacy may be violated, which may harm the user's benefits, both mentally and physically.  

To mitigate the potential risks and challenges in terms of ethics, related applications of recommender systems should adhere to data policies such as GDPR strictly, and give users the right to turn off the personal recommendation function.

\end{document}